\documentclass[fleqn,usenatbib]{mnras}

\usepackage{newtxtext,newtxmath}

\usepackage[T1]{fontenc}

\DeclareRobustCommand{\VAN}[3]{#2}
\let\VANthebibliography\thebibliography
\def\thebibliography{\DeclareRobustCommand{\VAN}[3]{##3}\VANthebibliography}

\usepackage{graphicx}	
\usepackage{amsmath}	
\usepackage{hyperref}
\usepackage{orcidlink}

\newcommand{\orcid}[1]{\href{https://orcid.org/#1}{\textcolor[HTML]{A6CE39}{\aiOrcid}}}

\title[Time-dependent AGN disks]{Time-dependent models of AGN disks with radiation from embedded stellar-mass black holes}

\author[Epstein-Martin, Tagawa, Haiman, \& Perna]{
Marguerite Epstein-Martin\orcidlink{0000-0001-9310-7808},$^{1}$\thanks{E-mail: mae2153@columbia.edu}
Hiromichi Tagawa,$^{2}$
Zolt\'an Haiman\orcidlink{0000-0003-3633-5403}$^{1,3}$
and Rosalba Perna\orcidlink{0000-0002-3635-5677}$^{4,5}$
\\
$^{1}$Department of Astronomy, Columbia University, 550 West 120th Street, New York, NY 10027, USA\\
$^{2}$Shanghai Astronomical Observatory, Shanghai, 200030, People$^{\prime}$s Republic of China\\
$^{3}$Department of Physics, Columbia University, 550 West 120th Street, New York, NY 10027, USA\\
$^{4}$Department of Physics and Astronomy, Stony Brook University, Stony Brook, NY 11794-3800, US\\
$^{5}$Center for Computational Astrophysics, Flatiron Institute, New York, NY 10010, USA
}

\date{Accepted XXX. Received YYY; in original form ZZZ}

\begin{document}
\maketitle

\begin{abstract}
The brightest steady sources of radiation in the universe, active galactic nuclei (AGN), are powered by gas accretion onto a central supermassive black hole (SMBH). The large sizes and accretion rates implicated in AGN accretion disks are expected to lead to gravitational instability and fragmentation, effectively cutting off mass inflow to the SMBH. Radiative feedback from disk-embedded stars has been invoked to yield marginally stable, steady-state solutions in the outer disks. Here, we examine the consequences of this star formation with a semi-analytical model in which stellar-mass black hole (sBH) remnants in the disk provide an additional source of stabilizing radiative feedback. Assuming star formation seeds the embedded sBH population, we model the time-evolving feedback from both stars and the growing population of accreting sBHs. We find that in the outer disk, the luminosity of the sBHs quickly dominates that of their parent stars. However, because sBHs consume less gas than stars to stabilize the disk, the presence of the sBHs enhances the mass flux to the inner disk. As a result, star formation persists over the lifetime of the AGN, damped in the outer disk, but amplified in a narrow ring in the inner disk.  Heating from the embedded sBHs significantly modifies the disk's temperature profile and hardens its spectral energy distribution, and direct emission from the sBHs adds a new hard X-ray component.\\
\end{abstract}



\section{Introduction} \label{sec:intro}

Active galactic nuclei (AGN) are widely understood to be the result of viscous accretion onto a central supermassive black hole (SMBH; $\gtrsim 10^6 \text{ M}_\odot$). Continuum spectra suggest that bright AGN are geometrically thin and optically thick \citep{shields1978,malkan1982}, conventionally modelled using the Shakura-Sunyaev $\alpha$ prescription \citep{shakura1973}. These `$\alpha$-disk' models are, however, well known to break down at large radii, becoming unstable to self-gravity beyond $\sim 0.1$ parsec (pc) and failing to transport mass quickly enough to maintain the disk beyond $\sim 1$ pc. In particular, the latter is expected to give rise to cascading fragmentation and star formation on a dynamical timescale -- rapidly transforming the disk into a flat stellar system and cutting off accretion onto the central SMBH \citep{schlosman1989, shlosman1990, goodman2003}. 

In an effort to resolve this apparent contradiction, \citet{collin1999} suggested that instability due to self-gravity may be mitigated by feedback from star formation including via stellar winds, supernovae explosions, and stellar accretion. Since then several models have been proposed which hinge on the assumption that fragmentation is self-limiting. In particular, \citet{sirko2003} and \citet[][hereafter TQM05]{TQM05} have
constructed modified viscous disk models in which the disk is assumed to maintain marginal stability at large radii. 

Though similar in their approach, these models are distinguished in their treatment of mass inflow through the AGN, resulting in
significant discrepancies in predicted disk density and scale height profiles. \citet{sirko2003} are strongly motivated by modeling the AGN infrared (IR) Spectral Energy Distributions (SEDs) and assume a constant rate of mass inflow to find the radiation pressure necessary to support the disk. In contrast TQM05 allow for changes in mass inflow, accounting for the star formation required to support the disk. Given the explicit relationship with star formation and more physically motivated mass dependencies, in the work that follows we will rely heavily on the original work of TQM05 (see Section 2.2 of \citet{fabj2020} and \citet{2024gangardt} for detailed comparisons).

The recent discoveries of gravitational waves (GWs) by LIGO-Virgo-KAGRA (LVK) have focused renewed attention on AGN disks, whose high density of embedded stars and compact objects make them an intriguing site for mergers and subsequent GW detections. Whether captured from the nuclear population \citep{bartos2017,panamarev2018, macleod2020, fabj2020, 1991syer} or formed in-situ \citep{stone2017}, once aligned with the disk, embedded objects are expected to exchange torque with the disk gas and migrate through the AGN. Migration can enhance the formation of new binaries via low-velocity encounters \citep{mckernan2012, leigh2018, tagawa2020} or upon entering annular gaps~\citep{tagawa2020} or migration traps \citep{bellovary2016, secunda2019,yang2019, secunda2020,Grishin+2024}, whereupon gas dynamical friction and binary-single encounters
act to harden binaries and promote mergers \citep{baruteau2011,tagawa2020,Tagawa2021_ecc,Tagawa2021_massgap, li2021, li2022, 2023DeLaurentiis, 2023boekholt, 2024Li}. 

Having shown that AGN disks can significantly alter the dynamics of stellar and compact object populations, it is pertinent to ask how such disks in turn may be affected by the objects they harbor. \citet{levin2003, levin2007} concluded that if BHs accreted at super-Eddington Bondi-limited rates, their luminosity would be sufficient to stabilize the AGN. However, embedded black holes accreting at highly super-Eddington rates for extended periods will quickly grow to intermediate masses, disrupting and depleting the AGN disk interior \citep{goodman2004, mckernan2012, stone2017}. Several feedback mechanisms have been suggested that could slow this accretion process \citep{inayoshi2016b, tagawa2022}. 
\citet{gilbaum2022} have recently developed a detailed model in which they assume pressure support from remnant stellar mass black holes (sBHs) fully overtakes the stellar component within one generation of massive star formation. They then calculate the effect of sBH radiation pressure support on a marginally stable, steady-state disk.  

In this paper, we develop a semi-analytic model for time-evolving AGN disks under the concurrent effects of star formation and sBH accretion, with the goal of understanding how a growing population of remnants impacts the disk structure, stellar distribution and electromagnetic spectrum of the disk over time. Assuming star formation seeds the population of sBHs, in \S\ref{sec:Equations} we introduce a set of equations to describe the evolution of the disk together with the embedded stars and sBHs. We use these equations in \S\ref{sec:Parameters} to discuss several timescales relevant to this model and define the parameters determining the number and mass distribution of disk-born stars. In \S\ref{sec:Feedback} we present order-of-magnitude estimates for the relation between mass accretion and radiation pressure from the two distinct populations of stars and sBHs. The numerical approach we used is explained in \S\ref{sec:implementation}. Our models build in complexity, first comparing AGN disks supported either by star formation or by sBHs alone (\S\ref{sec:steadystate}) and their respective spectral signatures. Then, in \S\ref{sec:evolution}, we model the response of the AGN disk to the combined effects of star formation and a growing populations of sBHs, approximating the evolution as a sequence of steady-states. We use these disk models to construct time evolving SEDs. In \S\ref{sec:HighMass} we examine the dependence of our models on disk viscosity and SMBH mass. In \S\ref{sec:Conclusions} we summarize our main conclusions and note avenues for future study. 


\section{Numerical model overview}
\label{sec:Equations}

We constructed our model in the steady-state Shakura-Sunyaev mold \citep{shakura1973}, assuming a geometrically thin, axisymmetric disk in local thermal equilibrium (LTE), undergoing quasi-viscous transport. As in TQM05, our model is distinguished from the standard $\alpha$-disk by requiring gravitational stability in the outer disk. The stability of a disk against small overdensities is typically represented by the dimensionless Toomre parameter $Q_T$ \citep{toomre1964}, where $Q_{\rm{T}} \lesssim 1$ indicates instability. Maintaining marginal stability requires

\begin{equation}
    Q_{\rm{T}} \simeq\frac{c_{\rm{s}} \Omega}{\pi G \Sigma} = \frac{\Omega^2}{\sqrt{2} \pi G \rho}\geq 1
\,,\label{eqn:Q}\end{equation}
where $c_{\rm{s}}$ is the sound speed related to the scale height $h$ by $c_{\rm{s}} = h \Omega$. $\Sigma = 2 \rho h$ is the surface density, $\rho$ is the disk density, 
$G$ is the gravitational constant, and $\Omega$ is the orbital frequency, given by
\begin{equation}
   \Omega = \left( \frac{G M_{\rm{sBH}}}{r^3} + \frac{2 \sigma^2}{r^2} \right)^{1/2}
\,.\label{eqn:Omega}\end{equation}
Here, $r$ is the distance from the SMBH, and 
$\sigma$ is the velocity dispersion characterizing the gravitational potential on galactic scales, beyond the central SMBH's sphere of influence.
In general, $\sigma$ scales with the mass of the central SMBH as $(\sigma/200\text{ km/s})^4 \sim M_{\rm{sBH}}/(2 \times 10^8\text{ M}_{\odot})$ \citep{2013kormandy}. Note that on parsec to tens of parsec scales, and for SMBH masses $10^6 - 10^9\text{ M}_{\odot}$, the second term in Equation (\ref{eqn:Omega}) is of the same order of magnitude as the standard Keplerian frequency. But in the interior disk the effect of the velocity dispersion term is negligible. 

From Equation (\ref{eqn:Q}) it follows that the density profile of the outer disk, where $Q=1$ is imposed, is solely determined by the orbital frequency. Moreover, from the Shakura-Sunyaev $\alpha$ viscosity parameter $\nu = \alpha c_s h$ and the continuity equation, the relationship between density and scale height may be calculated in terms of the mass accretion rate, i.e. 
\begin{equation}
   \dot{M} = 2\pi \nu \Sigma \left| \frac{d \ln \Omega}{d \ln r} \right| = 4 \pi G \alpha \Omega \rho h^3 \left| \frac{d \ln \Omega}{d \ln r} \right|
\,.\label{eqn:Mdot}\end{equation}

Equations (\ref{eqn:Q}-\ref{eqn:Mdot}) are identical to those presented in Appendix C of TQM05. Our model for the disk differs by the addition of new heating and mass accretion terms representing feedback from \textit{both} progenitor stars and their remnant black holes. In this disk, gas is consumed by star formation ($\dot{\Sigma}_{\star}$) and sBH accretion ($\dot{\Sigma}_{\text{BH}}$), so that the disk accretion rate decreases inward as
\begin{equation}
    \dot{M}(r) = \dot{M}_{\text{out}} - \int_r^{R_{\text{out}}} 2 \pi r (\dot{\Sigma}_{\star} + \dot{\Sigma}_{\text{BH}}) dr
\,,\label{eqn:Mdot2}\end{equation}
where $\dot{M}_{\text{out}}$ is the mass supplied at the outermost radius of the disk ($R_{\text{out}}$). 

In our models, marginal stability is maintained by vertical pressure support from star formation and sBH accretion. Following TQM05, we calculate the pressure associated with star formation as
\begin{equation}
    p_{\star} = \dot{\Sigma}_{\star}c\epsilon_{\star}\left(\frac{\tau}{2} +\xi\right)
\,,\label{eqn:pstar}\end{equation}
representing two distinct components: the radiation pressure on dust grains in the optically thick limit and the UV radiation pressure and turbulent support by supernovae in the optically thin limit. Here, $\epsilon_{\star}$ is the matter-radiation conversion efficiency of stars in the disk, discussed in more detail in \S\ref{subsec:Starformation}. The kinetic pressure associated with star formation is parameterized by the non-dimensional $\xi$, the ratio of star formation to supernovae pressure, which we set to $\simeq 1$ as in TQM05. 

Radiation pressure scales with the optical depth ($\tau = \kappa \rho h$) where $\kappa$ is the opacity. We calculate opacity according to \citet{semenov2003}.\footnote{The \citet{semenov2003} opacity tables do not include graphite grain opacity as discussed in \citet{2018baskin}, which dominates for hot dust. This additional component increases the opacity below 2000 K, more tightly constraining the opacity gap. However, we do not expect this adjustment to qualitatively change the conclusions made here and defer its inclusion to future work.} Because these opacity tables only extend to $10^4$ K, we follow the approach of TQM05, and smoothly connect the opacity to power laws given by \citet{bell1994} for temperatures exceeding $10^4$ K. Note that at temperatures of $\approx 1500$ K dust sublimation results in a steep opacity drop. The opacity rises again at $\approx 10^4$ K, with the ionization of hydrogen, creating a feature known as the `opacity gap' between $10^3-10^4$ K. This drop in opacity requires us to consider both the optically thick ($\tau \gg 1$) and thin ($\tau \ll 1$) regimes, in which the temperature varies with opacity as $T^4 \sim \tau T_{\rm{eff}}^4$ and $T^4 \sim \tau^{-1} T_{\rm{eff}}^4$, respectively. Interpolating between these two regimes, the temperature is
 \begin{equation}
     T^4 = T_{\rm{eff}}^4 \left(\frac{3}{4} \tau + \frac{1}{2 \tau} + 1\right)
 \,.\label{eqn:T}\end{equation}
Here, the effective temperature $T_{\rm{eff}}$ is calculated assuming thermal equilibrium in the disk,
\begin{equation}
    \sigma_{\rm{SB}} T_{\rm{eff}}^4 = \frac{3}{8 \pi}\dot{M}\left(1 - \sqrt{\frac{R_{\rm{in}}}{r}}\right)\Omega^2 + \frac{\dot{\Sigma}_{\star}\epsilon_{\star}c^2}{2} + \frac{Q_{\rm{sBH}}}{2}
\,,\label{eqnTeff}\end{equation}
where $\sigma_{\rm{SB}}$ is the Stefan-Boltzmann constant, $Q_{\rm{sBH}}$ represents the heating rate
associated with sBH accretion, and $R_{\rm{in}}$ is the innermost radius of the disk. Using Equations (\ref{eqn:pstar}) and (\ref{eqnTeff}), the total disk pressure can be written as
\begin{equation}
p_{\rm{tot}} = \rho c_{\rm{s}}^2 = p_{\rm{gas}} + \frac{\tau}{c}\sigma_{\rm{SB}}T_{\rm{eff}}^4 + \dot{\Sigma}_{\star}\epsilon_{\star} c\xi
\,,\label{eqn:ptot}\end{equation}
where $p_{\rm{gas}} = \rho k_{\rm{B}} T /  (\mu m_{\rm{p}} )$, $k_{\rm{B}}$ is the Boltzmann constant, $\mu$ is the gas mean molecular weight, and $m_{p}$ is the proton mass. Here, we set $\mu = 1.23$, appropriate for neutral gas with a primordial He/H mass ratio.

Having introduced $Q_{\rm{sBH}}$, the sBH heating term, we require at least one more equation to properly close our model. If we assume the number density of sBHs in the disk at a given time and radius is determined by the accumulated star formation, a self-consistent solution can be reached. We begin with an initial mass function (IMF) of the form $dN_{\star}/dm_{\star} = A_{\rm{x}} m_{\star}^{-\delta}$. Taking the mass of stars per unit disk surface area formed up to time $t$ to be $\int \dot{\Sigma}_{\star}(t) dt$ we can calculate the proportionality constant $A_{\rm{x}}$: 
\begin{equation}
    A_{\rm{x}}(t) = \frac{\int_0^t \dot{\Sigma}_{\star}(t) dt}{ \int_{m_{\rm{min}}}^{m_{\rm{max}}} m_{\star}^{1-\delta}dm_{\star}}
\,.\label{eqn:ax}
\end{equation}
The minimum ($m_{\rm{min}}$) and maximum ($m_{\rm{max}}$) stellar mass are free parameters which we take to be $0.1$ and $120 \text{ M}_\odot$, respectively. 

The number of remnant black holes can be found by taking the lower bound of the IMF and setting it to the turn-off mass ($m_{\rm{TO}}$) or the time-dependent mass at which stars are expected to evolve off the main sequence (MS). For ease of use we approximate the turn-off mass by a piecewise function, 
\begin{equation}
    m_{\rm{TO}} = \left\{
        \begin{array}{ll}
            m_{\rm{max}} & \quad t \leq t_{1} \\
            120\times10^{-\sqrt{\left(\log_{10}(t/\rm{yr}) - 6.43\right)/0.825}}& \quad t_{1}<t<t_{2}\\
            m_{\rm{trans}} & \quad t \geq t_{2}
        \end{array}
    \right.
\,.\label{eqn:mTO}
\end{equation}
Here, $m_{\rm{trans}} \simeq 20 \text{ M}_{\odot}$ is the transition progenitor mass between neutron stars (NS) and sBHs. $t_1$ and $t_2$ represent the time for the first and last sBH to be produced. That is, $t_{2}\simeq 10.0$ Myr is the time at which the smallest sBH progenitors evolve off the MS, and $t_{1}\simeq 2.7$ Myr is the minimum time required for the first sBHs to form. Between $t_{1}$ and $t_{2}$ we use the fitting function for $m_{\rm{TO}}$ given by Equation (3) of \citet{buzzoni2002}. 
The number of sBHs per unit disk area  at time $t$ can thus be written,
\begin{equation}
    S_{\rm{sBH}}(t)=\int_{t_0}^t A_{\rm{x}}(t')\int_{m_{\rm{TO}}(t - t')}^{m_{\rm{max}}}m_{\star}^{-\delta}dm_{\star} dt'
\,.\label{eqn:sbh}\end{equation}

With the number of sBHs in hand, calculating the corresponding heating and mass accretion terms is relatively straightforward. We assume an accretion rate onto embedded sBHs given by
\begin{equation}
\dot{m}_{\rm{sBH}} = \min\left(\dot{\rm{M}}_{\rm{Edd}}, \dot{m}_{\rm{B}}\right)
\,,\label{eqn:sBH accretion}
\end{equation}
i.e. capped at the Eddington rate:
\begin{equation}
\dot{\rm{M}}_{\rm{Edd}} = \frac{4 \pi G m_{\rm{sBH}}\mu_{\rm{e}} m_{\rm{p}}}{c \eta \sigma_{\rm{T}}}
\,.\end{equation}
Here $\mu_{\rm{e}} \sim 1.15$ is the mean weight per electron, $\sigma_{\rm{T}}$ is the Thompson cross section, $m_{\rm{p}}$ is the proton weight, and $\eta = 0.1$ is the radiative efficiency.

At sufficiently low density or scale height, the gravitational sphere of influence of an embedded sBH will not contain enough gas to sustain Eddington accretion, and the accretion will instead be set by the Bondi rate:
\begin{equation}
\dot{m}_{\rm{B}} = \pi r_{\rm{w}} r_{\rm{h}} \rho c_{\rm{s}}
\,,\label{eqn:mdotb}\end{equation}
where $r_{\rm{w}}$ and $r_{\rm{h}}$ are the width and height of the cross section of accretion. Oriented parallel to the plane of the disk, the width is calculated as
\begin{equation}
    r_{\rm{w}} = \min\left(R_{\rm{B}}, R_{\rm{Hill}}\right)
\,,\end{equation}
where $R_{\rm{Hill}} = r (m_{\rm{sBH}}/M)^{1/3}$ is the Hill radius \citep{1999murray} and $R_{\rm{B}} = G m_{\rm{sBH}} / c_{\rm{s}}^2$ is the Bondi radius \citep{bondi1952}. Perpendicular to the plane of the disk, $r_{\rm{h}} = \min(r_{\rm{w}}, h)$, ensuring that the cross sectional area does not extend above the vertical height of the disk \citep{stone2017, 2020rosenthal,2021dittmann}. 

Within the SMBH's sphere of influence
$\Omega \simeq \sqrt{G M/ r^3}$, and with some algebraic manipulation, we can re-define the three length scales -- $R_{\rm{B}}$, $R_{\rm{Hill}}$, and $h$ -- relative to one another: $R_{\rm{B}} > R_{\rm{Hill}}$ if $h < \sqrt{3} R_{\rm{Hill}}$. In terms of units relevant to this work, for an SMBH mass of $4\times10^6$ $M_{\odot}$ and sBH mass of 10 $\rm{M}_\odot$,  $\dot{m}_{\rm{B}}\propto R_{\rm{B}}^2$ where $h/r > 1.6 \times 10^{-2}$. In practice, this condition is nearly always satisfied.

Having specified the accretion rate for individual sBHs, the total sBH accretion per unit disk area can be written as 
\begin{equation}
\dot{\Sigma}_{\rm{sBH}} = \sum_i S_{\rm{sBH}}(m_{\rm{sBH}_i}) \dot{m}_{\rm{sBH}},(m_{\rm{sBH}_i})
\label{SigmadotBH}\end{equation}
where we have summed the accretion rate per unit area for sBHs of mass $m_{\rm{sBH}_i}$. The heating term is similarly
\begin{equation}
    Q_{\rm{sBH}} = \sum_i 
    S_{\rm{sBH}}(m_{\rm{sBH}_i}) \dot{m}_{\rm{sBH}}(m_{\rm{sBH}_i})
    \eta(m_{\rm{sBH}_i})  c^2 = \dot{\Sigma}_{\rm{sBH}} \eta c^2
\,.\label{eqn:QBH}\end{equation}
Here we assume that the radiative efficiencies during Bondi-limited and Eddington-limited accretion phases are the same, and set it to $\eta(m_{\rm{sBH}_i}) =\eta=0.1$, for simplicity. \footnote{In our models, sBH accretion does not fall below $\sim 0.01 ~\dot{\rm{M}}_{\rm{Edd}}$, justifying this assumption. If this were not the case, our models would need to be amended to account for the lower radiative efficiencies expected in the advection-dominated regime.}

In the high-density environment expected in AGN, the Bondi rate generally results in super-Eddington accretion. In our models, this means accretion is nearly always Eddington-capped. By limiting accretion to the Eddington rate, as in previous works (i.e. \citealt{gilbaum2022, tagawa2020}), the sBHs are prevented from quickly growing to intermediate sizes and depleting the disk interior \citep{goodman2004, stone2017}. On the micro-scale, Eddington-limited accretion is motivated by significant radiation pressure acting on dust grains in high metalicity gas \citep{2019toyouchi} or by wind mass loss \citep{Blandford1999}. Note that the latter can further reduce the mass flux through the disk.

\section{Parameter Choices and Regimes of Interest}
\label{sec:Parameters}

Our fiducial model assumes parameters appropriate to the Galactic center, namely an SMBH mass $M_{\text{BH}} = 4 \times 10^6 \text{ M}_{\odot}$ and velocity dispersion $\sigma = 75 \text{ km/s}$. The outer edge of the disk is defined as the radius where the velocity dispersion equals the Keplerian velocity or $R_{\text{out}} = G M_{\text{BH}}\sigma^{-2} \simeq 3 \text{ pc}$. The inner edge of the disk is assumed to be at the innermost stable circular orbit (ISCO) for a non-rotating black hole or $R_{\text{in}} = 3 \left(2 G M_{\text{BH}} c^{-2}\right)\simeq 10^{-6} \text{ pc}$. Following TQM05, we increase opacity by a factor of 3 to account for super solar metallicity in the Galactic center. We expand our discussion to a higher mass AGN in \S\ref{sec:HighMass}, setting $M_{\text{BH}} = 10^9~{\rm M_\odot}$.Details of other relevant parameter choices are discussed below. 

\subsection{Relevant Timescales}
\label{subsec:Timescales}

\begin{figure}
\centering
 \includegraphics[width=3.4 in]{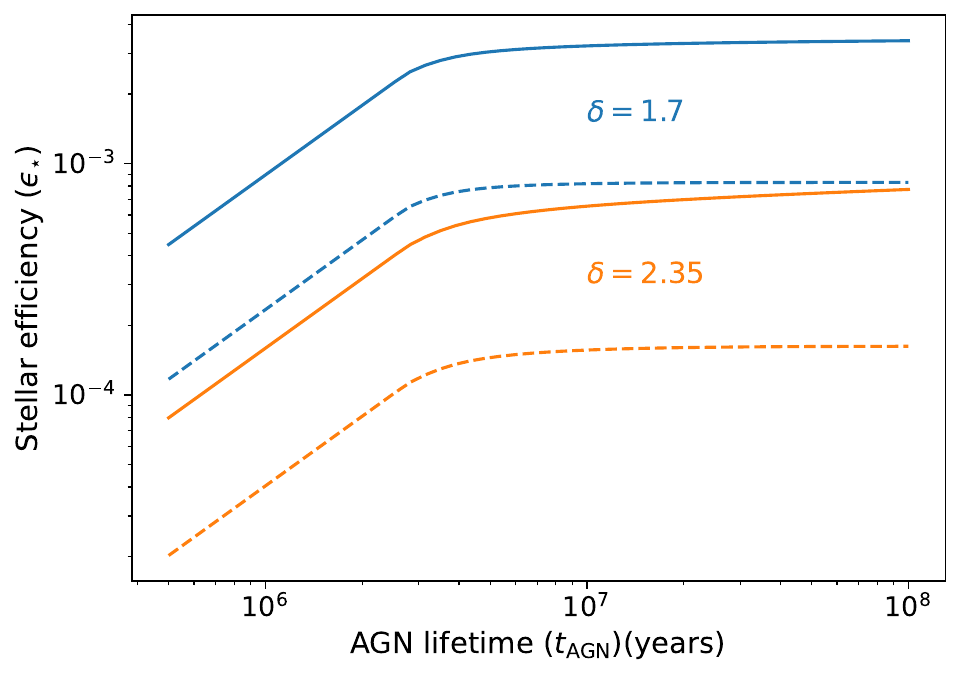}
 \caption{The efficiency of converting stellar mass to radiation calculated for the stellar IMF exponent $\delta_{\rm{IMF}} = 1.7$ (blue) and $\delta_{\rm{IMF}} = 2.35$ (orange). The solid lines illustrate the efficiency assuming the disk is optically thick to all wavelengths, while the dashed lines include only wavelengths in or above the UV or $\lambda \geq 100$ nm. Throughout this paper we assume a top heavy IMF ($\delta = 1.7$), an AGN lifetime of $10^8$ years, and a UV-limited stellar radiative efficiency of $\epsilon_\star = 8.3\times10^{-4}$ as defined by the dashed blue line. }
    \label{fig:efficiency}
\end{figure}

As mentioned in the Introduction, one problem for AGN disk models is the long viscous timescales predicted by the `$\alpha$'-viscosity parameterization, which suggests outer viscous timescales at least an order of magnitude longer than anticipated AGN lifetimes \citep{schlosman1989}. Most proposed resolutions of this problem invoke increased angular momentum transport efficiency via non-axisymmetric global-torques, which could be provided by stellar bars, spiral waves, or magnetic stresses \citep{shlosman1990, hopkins2011}. 
\citet{gilbaum2022} showed, however, that the luminosity from a growing population of sBHs may serve to increase the disk scale height, which scales inversely with the viscous timescale, partially alleviating the inflow problem. In order to similarly examine the relative change in viscous timescale we also assume a standard local $\alpha$-viscosity prescription. The kinematic viscosity may then be written as $\nu = \alpha c_{s} h$, with the dimensionless parameter $\alpha$ set to 0.3 in this work. 

The viscous timescale $t_{\nu}$ is also important in determining the critical mass supply rate ($\dot{M}_{\rm{c}}$), identified as the minimum $\dot{M}_{\rm{out}}$ below which star formation in the outer disk significantly depletes the disk mass. This limit corresponds to when the viscous timescale ($t_{\nu}$) is shorter than the star formation timescale or $t_{\rm{SF}} < t_\nu $ where $t_{\rm{SF}} = \Sigma/\dot{\Sigma}_{\star}$. By making a few physically motivated approximations at the outer disk we can use this inequality to calculate $\dot{M}_{\rm{c}}$. Namely, we assume  $Q_{\rm T} = 1$, and that the disk is optically thick and radiation-pressure dominated. In the outer disk dust opacity dominates which has the temperature dependence $\kappa = \kappa_0 T^2$. This yields a star-formation timescale of $t_{\rm{SF}} \propto \epsilon_{\star} \kappa_0 T^2$ and a viscous timescale $t_{\nu} \propto r^2 T^2/\dot{M} $. The critical mass supply rate is then expressed as (see also Equation~44 in TQM05 and the Appendix of \citealt{inayoshi2016}), 
\begin{equation}
\begin{split}
    \dot{M}_{\rm{c}} =  &0.26\, \left(\frac{r}{3 \text{ pc}}\right)^{2}
    \left(\frac{\kappa_0}{7.2 \times 10^{-4} {\rm cm^2\, g^{-1}\, K^{-2}}}\right)^{-1}
    \\&
    \times\left(\frac{\epsilon_{\star}}{8.3 \times 10^{-4} }\right)^{-1}  \left(\frac{M_{\rm{BH}}}{4\times10^6 ~\rm{M}_{\odot}}\right)^{-1}\dot{\rm{M}}_{\rm{Edd}}
\,. \label{eqn:mc}
\end{split}\end{equation}
Note that to find $\dot{M}_{\rm{c}}$ for a disk supported by sBHs, we replace $\epsilon_\star$ with $\eta = 0.1$ and find $\dot{M}_{\rm{c}}\simeq 10^{-3}~\dot{M}_{\rm{Edd}}$ -- far below the parameter space with which we are concerned here. In our disk models it can thus be safely assumed that $\Sigma/\dot{\Sigma}_{\rm{sBH}}\gg t_{\nu}$.

Another important timescale is the lifetime of the AGN disk ($t_{\rm{AGN}}$). The AGN duty cycle is still somewhat controversial, with expected lifetimes ranging from $10-10^3$ Myr \citep{martini2004}. In the simplest case, these lifetimes are consistent with the mass-independent, Eddington limited e-folding time for the SMBH, i.e. $t_{\rm{AGN}} = \eta \sigma_{\rm{T}} c/(4 \pi G m_{p}) \sim 50(\eta/0.1)$ Myr \citep{salpeter1964}. In the shortest AGN lifetime limit, massive sBH progenitors will not have time to form and leave the MS. Therefore, we discount these regimes focusing on AGN lifetimes exceeding $10^7$ years.

There are two additional physically relevant timescales: the thermal ($t_{\rm{th}}$) and the dynamical  ($t_{\rm{dyn}}$) time. The latter scales as $\Omega^{-1}$ and the thermal timescale, or the disk diffusion timescale, is given by $\Sigma c_{\rm{s}}^2/(\sigma_{\rm{b}} T_{\rm{eff}}^4) \simeq \Sigma c_{\rm{s}}^2/(\epsilon_\star \dot{\Sigma}_\star c^2/2) $. In the optically thick, radiation-pressure dominated limit, the ratio of the thermal and dynamical timescales is $t_{\rm{th}}/t_{\rm{dyn}} \sim \tau c_{\rm{s}}/c$. Because $\tau$ never exceeds 1000 in our models, we conclude that $t_{\rm{th}}/t_{\rm{dyn}}\ll1$. Where the disk is optically thin, $t_{\rm{SF}}\simeq t_{\nu}$. Substituting this into $t_{\rm{th}}$, we find $t_{\rm{th}}/t_{\rm{dyn}} \simeq 2 (\Omega r)^2/(\epsilon_\star \alpha c^2) \ll 1$.  This inequality implies that the disk should self-regulate, such that $Q\sim1$ as our model assumes. \footnote{We can similarly compare $t_{\rm{dyn}}$ and $t_{\rm{SF}}$, finding that $t_{\rm{SF}}/t_{\rm{dyn}} \sim \tau \epsilon_{\star} c_{\rm{s}}/c$ when $p_{\rm{rad}}\gg p_{\rm{gas}}$ and $\tau\gg1$. For $\epsilon_{\star}\ll 1$, $t_{\rm{dyn}}\ll t_{\rm{SF}}$.  In the optically thin limit, $t_{\rm{th}}/t_{\rm{SF}} = 2c_{\rm{s}}^2/(\epsilon_\star \alpha c^2)$, dividing this ratio by $t_{\rm{th}}/t_{\rm{dyn}}$ we find $t_{\rm{dyn}}/t_{\rm{SF}} \simeq (h/r)^2 \ll 1$. Thus we can express the relationship between timescales as $t_{\rm{th}}\ll t_{\rm{dyn}}\ll t_{\rm{SF}}$, and expect it to hold for all relevant disk parameters. Because $t_{\rm{th}}\ll t_{\rm{SF}}$, $\dot{\Sigma}_{\star}$ can be assumed constant on timescales of $t_{\rm{th}}$. This has implications for thermal disk stability, as noted in Appendix B.2 of TQM05.}

\subsection{Star formation}
\label{subsec:Starformation}

In determining the number and mass distribution of disk-born stars we must define several parameters including: the minimum (${m}_{\rm{min}}$) and maximum (${m}_{\rm{max}}$) stellar masses, the IMF power-law index $\delta$, and the efficiency of converting stellar mass to radiation ($\epsilon_{\star}$). In this work, we assume a minimum stellar mass of 0.1 $\text{M}_{\odot}$, the approximate mass required to ignite thermonuclear reactions in the stellar interior. The maximum mass of 120 $\text{M}_{\odot}$ is less rigorously defined, and is likely a lower limit \citep{2021cantiello, 2021dittmann, 2022jermyn}, but given $m_{\rm{min}}\ll m_{\rm{max}}$ and therefore $N_{\star} \sim A_{x} m_{\rm{min}}^{1-\delta}$, its value is unlikely to strongly effect the outcome of our models.

We consider two different IMF slopes: the standard Salpeter IMF $\delta = 2.35$ determined using the luminosity function of stars in the solar neighborhood \citep{salpeter1955} and a top-heavy IMF $\delta = 1.7$, an observational estimate based on the Milky Way's nuclear stellar disk \citep{lu2013}. These $\delta$ values also encompass the range of IMF slopes investigated in the AGN models of \citet{tagawa2020}. 

The stellar mass conversion efficiency has a range of values in the literature, likely because the calculation requires approximations of the lifetime of the AGN as well as stellar lifetimes and luminosities across the stellar mass spectrum. In the original work of TQM05, an efficiency of $\epsilon_\star=10^{-3}$ was assumed for an IMF slope $\delta = 2.35$, although the stellar lifetime was not considered in their calculation. More recently, \citet{tagawa2020} recalculated the efficiency including the effect of stellar and AGN lifetimes, calculating an efficiency of $1.5 \times 10^{-4}$ and $7.7 \times 10^{-4}$ for an IMF exponent $\delta = 2.35$ and $1.7$ respectively, an AGN lifetime of $t_{\rm{AGN}} = 10^8$ years, and stellar lifetime of $t_{\rm{star}} = 10$ Gyr $(m_\star/{\rm \rm{M}_\odot})(L_\star/{\rm L_\odot})^{-1}$. Here, we calculate the efficiency for the two adopted stellar IMFs scaling by stellar lifetime and an efficiency assuming UV radiation alone supports the AGN disk. This is likely to be the case in the outer disk where gas-coupled dust grains make the disk optically thin to re-radiated IR radiation but optically thick to UV radiation. Here, we compute the UV-weighted case and compare with the non-UV weighted efficiency assumed in previous works. Efficiency is calculated according to Equation~A4 from \citet{tagawa2020}, 
\begin{equation}
    \epsilon_{\star}= \frac{\int_{m_{\rm{min}}}^{m_{\rm{max}}}L(m_{\star}) \text{min}[t_{\rm{AGN}}, t_{\star}(m_{\star})]m_{\star}^{-\delta} dm_{\star}}{\int_{m_{\rm{min}}}^{m_{\rm{max}}}m_{\star}c^2m_{\star}^{-\delta}dm_{\star}}
\,,\label{eqn:efficiency}\end{equation}
assuming the same mass-stellar lifetime relation
\begin{equation}
t_{\star} = 10^{ 0.825\log^2({m_{\star}/120 {\rm M}_{\odot}})  + 6.43} \,{\rm yr}
\end{equation}
as in Equation (\ref{eqn:mTO}). The stellar mass-luminosity relation was taken to be a piecewise function (e.g. Equation~A2 in \citealt{tagawa2020} and \citealt{solaris2005}):
\begin{equation}
    L(m_{\star}) = \left\{
        \begin{array}{ll}
            0.27 \text{ L}_{\odot} (m_{\star}/\text{M}_{ \odot})^{2.6} & \quad m_{\star} < 0.5 \text{M }_{\odot} \\
           \text{ L}_{\odot} (m_{\star}/\text{M}_{\odot})^{4.5} & \quad 0.5 \text{ M}_{\odot}< m_{\star} < 2 \text{ M}_{\odot}\\
           1.9\text{ L}_{\odot} (m_{\star}/\text{M}_{\odot})^{3.6} & \quad 2 \text{ M}_{\odot}< m_{\star} < 42 \text{ M}_{\odot}\\
           32000\text{ L}_{\odot} (m_{\star}/\text{M}_{\odot})
           & \quad 42 \text{ M}_{\odot}< m_{\star} < 120 \text{ M}_{\odot}
        \end{array}
    \right.
\,.\label{eqn:luminosity}\end{equation}
Together, Equations (\ref{eqn:efficiency}) and (\ref{eqn:luminosity}) allow us to calculate the non-UV weighted stellar efficiency as a function of AGN lifetime, illustrated by the solid lines in Fig.~\ref{fig:efficiency}. In order to determine the UV-weighted efficiency, we assumed a stellar radius-mass relation $R_{\star} \propto m_{\star}^{1/2}$ and effective temperature scaling $T_{\rm{eff}}\propto (L_{\star}/R_{\star}^2)^{1/4}$. Assuming a black-body spectrum for stars we then calculated the stellar UV luminosity as 
\begin{equation}
    L_{\star_{\rm{UV}}}(m_{\star}) = \int_0^{100 \text{nm}} \frac{2 h_{\rm{p}} c^2/\lambda^5}{\exp\left[\frac{h_{\rm{p}} c}{\lambda k_{\rm{b}} T_{\rm{eff}}(m_{\star})}\right] - 1} d\lambda\,,
\label{eqn:LstarUV}\end{equation}
where $h_{\rm{p}}$ is the Plank constant, and 
our upper bound of 100 nm limits the computed luminosity to the UV range and above. Substituting Equation (\ref{eqn:LstarUV}) into (\ref{eqn:efficiency}) 
we calculated the UV-weighted stellar efficiency as a function of AGN lifetime, described by the dashed curves in Fig.~\ref{fig:efficiency}. For a Salpeter IMF, the UV and non-UV weighted efficiencies are $\epsilon_\star = 1.6 \times 10^{-4}$ and $7.7 \times 10^{-4}$ respectively, while for a top heavy IMF they  are $8.3 \times 10^{-4}$ (UV-weighted) and $3.4 \times 10^{-3}$ (non-UV weighted). The discrepancy between the non-UV weighted efficiencies calculated here and those calculated in \citet{tagawa2020} come from the difference in approach to stellar lifetime. For the remainder of this paper we assume a top-heavy IMF and a UV-weighted efficiency of $\epsilon_{\star} = 8.3 \times 10^{-4}$.

\section{Sources of Feedback}
\label{sec:Feedback}

It is illustrative to compare the anticipated values of luminosity and accretion rates for a single generation of star formation and a remnant sBH population. We follow the procedure laid out in Equations (\ref{eqn:ax}) and (\ref{eqn:sbh}), substituting $\dot{\Sigma}_{\star} \delta t$ into our equation for $A_{\rm{x}}$. We can then solve for the number of black holes per unit disk area, assuming the boundary of the integral taken at the transition mass between sBHs and NSs ($m_{\rm{trans}} \simeq 20 ~\rm{M}_\odot$) to include all sBHs produced from the mass of stars $\dot{\Sigma}_{\star}\delta t$. Solving for the relative stellar and sBH heating components we find, 
\begin{equation}
\begin{split}
    \frac{Q_{\rm{sBH}}}{Q_{\star}} =& \frac{S_{\rm{sBH}}\frac{4 \pi G \mu_{\rm{e}}m_{\rm{p}} m_{\rm{sBH}}c}{\sigma_{\rm{T}}}}{\dot{\Sigma}_{\star}\frac{\epsilon_{\star}c^2}{2}}\\&
    \simeq 1.9 
    \left(\frac{m_{\rm{sBH}}}{10 \text{ M}_{\odot}}\right)\left(\frac{\delta t}{3\text{ Myr}}\right)\left(\frac{\epsilon_\star}{8.3\times10^{-4}}\right)^{-1}\,,
\end{split}
\label{eqn:stellar_sBH_heating}\end{equation}
where we have assumed a top heavy IMF of $\delta = 1.7$. This suggests that heating from sBHs should overtake stars within a single generation or approximately the time it takes for the most massive stars to evolve off the main sequence ($\simeq 3$ Myr).

Under the assumption that the sBH luminosity will quickly surpass stellar luminosity in AGN, \citet{gilbaum2022} did not include radiation pressure or disk mass depletion due to star formation in their models. However, in comparing the mass accretion rates of these two populations we found, 
\begin{equation}
    \frac{\dot{\Sigma}_{\rm{sBH}}}{\dot{\Sigma}_{\star}} = \frac{\epsilon_{\star}}{\eta}\frac{Q_{\rm{sBH}}}{Q_{\star}}
\,,\label{eqn:stellar_sBH_accretion}\end{equation}
indicating that $\dot{\Sigma}_{\rm{sBH}} \ll \dot{\Sigma}_{\star}$ for expected 
sBH radiative efficiencies between $0.1 \leq \eta \leq 1$. The significant disparity in the mass accretion rates between stars and sBHs suggests that as radiation pressure shifts from being star-formation to sBH-dominated, the mass flow through the disk should increase, since less mass is removed by sBH accretion than by star formation, to achieve the same heating rate. The luminosity from the sBHs in the disk interior is set by the preceding generation of star formation. If this sBH luminosity can not sustain disk stability in the face of increased mass flux, additional star formation will be necessary to stabilize the disk interior. 

 Note that in our discussion of feedback mechanisms we have neglected feedback from neutron stars (NSs) and white dwarfs (WDs). We discount these sources on the basis that NS masses and therefore Eddington-limited accretion rates are expected to be an order of magnitude lower than their sBH counterparts. WD feedback is also ignored under the assumption that a significant population would not have time to form within a typical AGN lifetime.

\subsection{X-ray opacity}
\label{subsec:XrayOpacityn}

\begin{figure}
\centering
 \includegraphics[width=3.2 in]{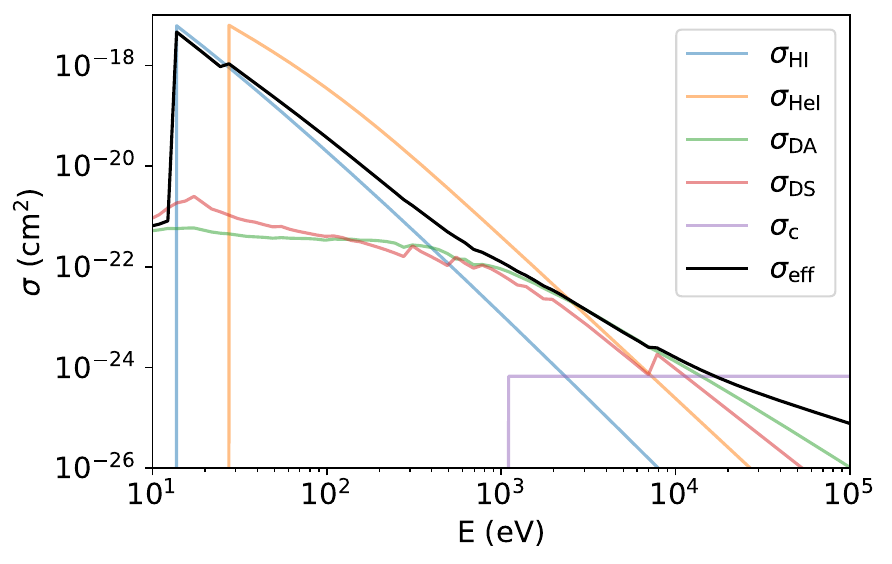}
 \caption{Cross-sections as a function of photon energy for photoionization of hydrogen (blue) $\sigma_{\rm HI}$ and singly ionized helium (orange) $\sigma_{\rm HeI}$. The cross-section for dust absorption $\sigma_{\rm{DA}}$ and scattering $\sigma_{\rm{DS}}$ are plotted in green and red respectively, while in purple we display the Compton cross-section for neutral hydrogen, $\sigma_{\rm{C}} = \sigma_{\rm{T}}$ when $E > 1$ keV. The effective cross-section $\sigma_{\rm{eff}}$ (black) is defined in Equation (\ref{eqn:sigma_eff}) in the text. This parameter is used to determine the disk's optical depth to X-rays and the relative fraction of sBH X-ray emission we expect to escape the disk.}
    \label{fig:xsection}
\end{figure}

In TQM05 it is argued that UV radiation from massive stars is absorbed and scattered by dust grains, which reprocess the UV emission into the IR. As applied in TQM05, $\tau$ refers to the optical depth of the disk to IR photons and not to the stellar UV emission to which the disk is always optically thick. Moreover, the temperature profiles presented in TQM05 do not exceed $10^4$ K in the star-forming, outer disk. At such low temperatures, we expect the disk gas to be neutral, interacting with the emitted UV radiation via photoionization. This setup is not unlike models for feedback in the inter-stellar medium (ISM) (e.g. \citealt{1977mckee}), in which stellar radiation creates hot ionized bubbles in a cold inter-bubble medium. These bubbles emit radiation from recombination in the IR. In the AGN disk context, radiative diffusion ensures that the momentum from these re-radiated IR photons is efficiently coupled to the gas and uniformly heats the disk. 

We expect the X-ray flux emitted from accreting sBHs to interact with dust and gas in the disk in a way similar to their parent stars. That is, dust should scatter and absorb X-rays while neutral gas is photoionized. Additionally, high energy X-rays (> 1 keV) undergo Compton scattering by the electrons bound in neutral hydrogen and helium atoms~\citep{1996sunyaev}. Still, because the mean free path of X-ray photons is large relative to UV photons, we cannot assume the disk is always opaque to X-rays. In this section, we calculate the optical depth of X-rays in the disk and determine $f_{\rm{esc}}$, the fraction of sBH emission we anticipate escaping the disk. 

We begin by calculating an effective cross-section for absorption and scattering of X-ray photons according to \citep{1986rybicki},
\begin{equation}
    \sigma_{\rm{eff}} = \sigma_{\rm{abs}} \sqrt{1 + \frac{\sigma_{\rm{scat}}}{\sigma_{\rm{abs}}}}\,.
\label{eqn:sigma_eff}\end{equation}
Here, $\sigma_{\rm{abs}}$ and $\sigma_{\rm{scat}}$ are the cross-sections for X-ray absorption and scattering respectively. Included in $\sigma_{\rm{abs}}$ is dust absorption and photoionization of hydrogen and helium,
\begin{equation}
\sigma_{\rm{abs}} = X \left(\sigma_{\rm{HI}} + \sigma_{\rm{DA}}\right) + \frac{Y}{4} \sigma_{\rm{HeI}}
\end{equation}
where $\sigma_{\rm{DA}}$ is the dust absorption taken from \citet{2003draine} and $X = 0.75$ and $Y = 0.25$ are the hydrogen and helium mass fractions respectively, assuming a primordial gas composition. The equation for the photoionization cross section of hydrogen $\sigma_{\text{H I}}$ is adopted from Equation (2.4) of \citet{2006osterbrock},
\begin{equation}
\sigma_{\text{HI}} = A_0 \left( \frac{E_1}{E}\right)^4 \frac{\exp{\left(4 - 4 \tan^{-1} \left(\mathcal{E}\right)/\mathcal{E}\right)}}{1 - \exp\left(-2\pi\mathcal{E}\right)}
\,,\label{eqn:crossSection}
\end{equation}
where $A_0 = 6.30 \times 10^{-18} \text{ cm}^2$,
$\mathcal{E} \equiv (E/E_1 - 1)^{1/2}$,
$E_1 = h_P\nu_1  = 13.6 $ eV is the hydrogen ionization threshold energy, and $h_P$ is Planck's constant. From \citet{1996Haardt}, we calculate the cross section for singly ionized helium as,
 \begin{equation}
     \sigma_{\text{HeI}} \simeq \frac{0.694 \times 10^{-18}}{(E/10^2 \text{ eV})^{1.82} + (E/10^2\text{ eV})^{3.23}} \text{ cm}^2\,.
\label{eqn:HeI} \end{equation}

The scattering cross-section $\sigma_{\rm{scat}}$ is taken to be the sum of dust and Compton scattering from neutral atoms. The dust scattering cross-section per hydrogen atom $\sigma_{\rm{DS}}$ is taken from \citet{2003draine}, for Milky-Way dust.
The Compton cross-section for neutral hydrogen $\sigma_{\rm{C}}$ is set equal to the Thompson cross-section ($\sigma_T = 6.65 \times 10^{-25}\text{ cm}^2$) for photon energy $E > 1$ keV, and zero everywhere else. Following \citet{1996sunyaev}, we multiply $\sigma_{\rm{C}}$ by 1.5 to account for additional scattering by helium and other metals, allowing us to write the scattering cross-section as 
\begin{equation}
    \sigma_{\rm{scat}} = X \left(1.5\sigma_{\rm{C}} + \sigma_{\rm{DS}}\right)\,.
\end{equation}
The effective cross-section $\sigma_{\rm{eff}}$ and component cross-sections are plotted in Fig.~\ref{fig:xsection}. 

 \begin{figure}
\centering
 \includegraphics[width=3.2 in]{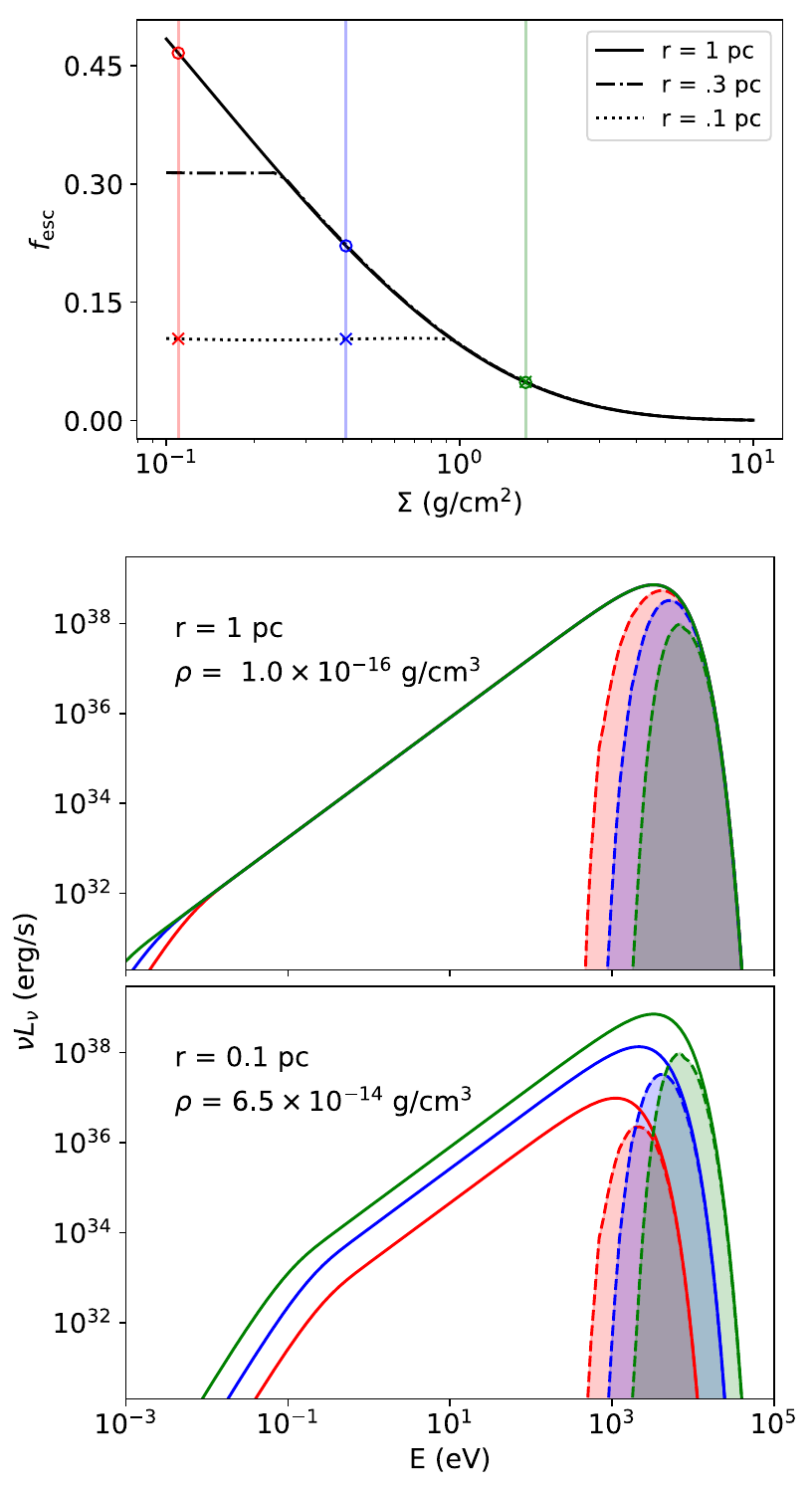}
 \caption{{\em Top panel}: The escape fraction of ionizing radiation as a function of $\Sigma$, for a 10 $\rm{M}_\odot$ sBH embedded at a distance of $r=1$ (solid black line), $r=0.3$ pc (dot dashed line) and $r = 0.1$ pc (dotted line) from the central SMBH. We adopt the volumetric density of a marginally stable disk, i.e. $\rho = \Omega^2/(\sqrt{2}\pi G)$. A red, blue and green circle with coordinates $\Sigma({\rm g/cm^2}) = 0.11$, $0.41$, and $1.68$ and $f_{\rm{esc}} = 0.47$, 0.22, and 0.05 respectively, are plotted along the $r=1$~pc curve. Blue, red, and green crosses mark the same $\Sigma$ values along the $r=0.1$ pc curve, with $f_{\rm{esc}} = 0.10$, 0.10, and 0.05. The circles and crosses correspond to the spectra in the same colors in the middle and bottom panels respectively. Note that at 1 pc, sBH accretion is Eddington-limited across plotted values of $\Sigma$. In the Eddington-limited regime, $f_{\rm{esc}}$ scales inversely with $\Sigma$, otherwise $f_{\rm{esc}}$ is approximately constant. {\em Middle panel}: Three example spectra for a 10 $\rm{M}_\odot$ sBH embedded at $r=1$ pc. Solid lines indicate the intrinsic (unabsorbed) sBH emission spectra. Filled in regions illustrate the radiation that escapes from the disk. {\em Bottom panel}: Same as the middle panel but for $r=0.1$ pc.}
    \label{fig:sBH_spectra}
\end{figure}

An X-ray photon will escape the gas disk if the effective optical depth of the disk to X-rays is $\tau_{\rm{X}} < 1$. Writing $\tau_{\rm{X}}$ as the ratio of the half-thickness $h$ 
and the X-ray mean free path $\Lambda_{\rm{X}}$, the fraction of radiation from a single sBH to escape the disk is calculated as
\begin{equation}
f_{\rm{esc}} = \frac{\int_{\nu_1}^{\infty} L_{\rm{sBH}_\nu}\exp{\left(-\frac{h}{\Lambda_{\rm{X}}}\right)}d\nu} 
{ \int_{\nu_1}^{\infty} L_{\rm{sBH}_\nu} d\nu}
\,.\label{eqn:fesc}\end{equation}
Here, the mean free path is defined in terms of the effective cross section and disk density as $\Lambda_{\rm{X}} = (\rho/m_{\rm{p}} \sigma_{\rm{eff}})^{-1}$ and emission from an accreting sBH ($L_{\rm{sBH}_\nu}$) is modeled as a Shakura-Sunyaev accretion disk, with a mass flux $\dot{m}_{\rm{sBH}}$, inner radius equal to the innermost stable circular orbit ($r_{\rm{ISCO}} = 3 R_{\rm{s}}$), and outer radius of $r_{\rm{h}}$.  

The top panel of Fig.~\ref{fig:sBH_spectra} shows $f_{\rm{esc}}$ as a function of $\Sigma$ for a $10~{\rm \rm{M}_\odot}$ sBH embedded in the disk at a radius of 1 pc (solid black line), 0.3 pc (dot-dashed line) and 0.1 pc (dotted line). We assume a marginally stable disk with volumetric density $\rho = \Omega^2/(\sqrt{2}\pi G)$, such that $\rho = 1.0 \times 10^{-16}$, $2.7 \times 10^{-15}$, and $6.5 \times 10^{-14} \text{g/cm}^3$ at $r = 1$, 0.3, and 0.1 pc respectively. 

When sBH accretion is Eddington capped, $f_{\rm{esc}}$ is independent of the location of the embedded sBH in the AGN disk and follows the solid $r = 1$ pc curve. In the top panel of Fig. \ref{fig:sBH_spectra}, red, blue, and green circles plotted along the along the solid line correspond to the same color sample spectra in the middle panel, whose shaded regions illustrate the radiant energy that escapes the disk. Note that the Eddington-capped intrinsic sBH spectra—shown as solid curves in the middle panel of Fig. \ref{fig:sBH_spectra}—are largely independent of disk parameters. Small differences between spectra at the lowest energies arise because the sBH accretion disks are truncated at $r_{\rm{h}} = h$. In these cases, the dependence of $f_{\rm{esc}}$ on the local disk parameters is limited to the exponential term in Equation (\ref{eqn:fesc}), which we can re-write in terms of $\Sigma$ explicitly: $f_{\rm{esc}} \propto \exp(-\Sigma/m_{\rm{p}} \sigma_{\rm{eff}})$.

Closer to the SMBH, at $r = 0.1$ pc (0.3 pc), $f_{\rm esc}$ diverges from the solid curve and flattens for $\Sigma \lesssim 1 ~\rm{g/cm}^2$ ($0.3 ~\rm{g/cm}^2$) dropping to an approximately constant value of $f_{\rm{esc}} \simeq 0.1$ ($0.3$) as indicated by the dotted (dash-dotted) line in the top panel of Fig. \ref{fig:sBH_spectra}. Red, blue, and green crosses on the $r = 0.1$ pc curve correspond to the same color spectra in the bottom panel of Fig. \ref{fig:sBH_spectra}. The escape fraction drops between $r = 1$ and 0.1 pc for $\Sigma \lesssim 1 ~\rm{g/cm}^2$ because $\dot{m}_{\rm{sBH}}$ is not Eddington capped; instead, $\dot{m}_{\rm{sBH}} = \dot{m}_{\rm{B}}$, determined by $r_{\rm{w}} = R_{\rm{Hill}}$ and $r_{\rm{h}} = h$, with $\dot{m}_{\rm{sBH}} \propto \Sigma^2$ (Equation (\ref{eqn:mdotb})). The decrease in $\dot{m}_{\rm{sBH}}$ lowers both the energy and amplitude of the intrinsic sBH spectra, as seen in the red and blue solid curves plotted in the bottom panel. Despite the reduced sBH emission, $f_{\rm{esc}}$ remains approximately constant for $\Sigma \lesssim  1 ~\rm{g/cm}^2$ because $\tau_{\rm{X}} \propto \Sigma$. For $\Sigma = 1.68 ~\rm{g/cm}^2$ (green), $f_{\rm{esc}} = 0.05$ at $r = 0.1$ and 1 pc, as $\dot{m}_{\rm{sBH}} = \dot{\rm{M}}_{\rm{Edd}}$ in both cases, resulting in identical spectra above 1 eV.

Note that in a marginally stable disk the volumetric density scales with radius as $\rho \propto r^{-3}$, and the scale height goes as $h \propto (\dot{M}(\rho\Omega)^{-1})^{1/3} = \dot{M}^{1/3}r^{3/2}$. The column density is therefore proportional to $\Sigma \propto \dot{M}^{1/3}r^{-3/2}$ and we expect $\Sigma$ to scale inversely with $r$. We verify this general trend in our steady state models, discussed in \S\ref{sec:steadystate}, plotting $f_{\rm{esc}}$ as a function of distance from the SMBH in the top panel of Fig. \ref{fig:f_esc_L_esc}. Note too the second row of Fig. ~\ref{fig:steadyState} where we plot $\Sigma$. We find that where the disk is marginally stable, and changes in $\dot{M}$ are small, our scaling relation holds. This plot also shows that in general $\Sigma \sim 10 ~(\rm{g/cm}^3)$ at $r\sim 0.1$ pc while $\Sigma \sim 0.1 ~(\rm{g/cm}^3)$ at $r \sim 1$ pc. We therefore expect sBH accretion to be Eddington-limited across these models. 

Finally, in order to incorporate the reduction in the heating rate due to some of the radiation escaping from the disk, we amend Equation~(\ref{eqn:QBH})~to
\begin{equation}
Q_{\rm{sBH}} = S_{\rm{sBH}} \dot{\Sigma}_{\rm{sBH}} \eta c^2  \left(1 - f_{\rm{esc}}\right). \label{eqn:QBHNew}
\end{equation}
Note that the loss of sBH radiation is equivalent to a decrease in the effective radiative efficiency as it relates to disk heating, i.e. $(1-f_{\rm esc})$ is degenerate with $\eta$. 
The radiation loss decreases the effective heating rate of sBHs relative to the stellar disk component such that the ratio $Q_{\rm{sBH}}/ Q_{\star}$ (calculated in Equation~\ref{eqn:stellar_sBH_heating}) decreases and the time required until sBH radiation dominates stellar radiation increases.

\section{Numerical Implementation}
\label{sec:implementation}

\begin{figure*}
\centering
 \includegraphics[width=7.3 in]{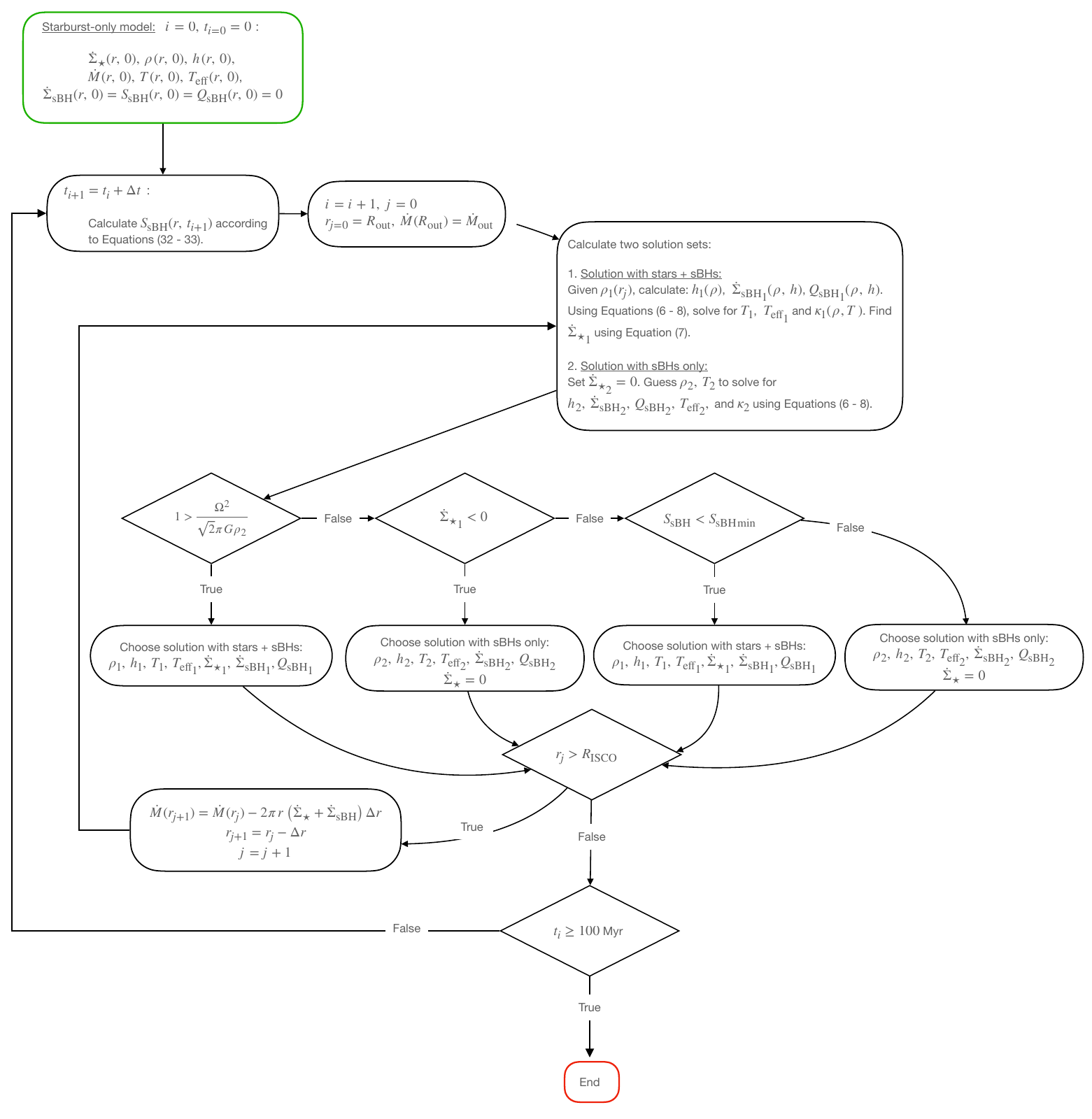}
 \caption{Schematic detailing our solution strategy in our disk evolution models. Beginning with a starburst only model at $t = 0$, we calculate the distribution of sBHs across the disk according to the preceding star formation rates. We then choose between marginally stable solutions including pressure support from star formation + sBH accretion and disk solutions supported by sBHs only (differentiated by subscript 1 or 2 in the schematic). The chosen solution gives $\rho$, $h$, $T$, $T_{\rm{eff}}$, $\dot{\Sigma}_\star$, and $\dot{Q}_{\rm{sBH}}$ for each radius starting from $R_{\rm{out}}$ and moving inward. This procedure is repeated at each time step until $t = 100$ Myr.}
    \label{fig:implementation}
\end{figure*}

The equations outlined in \S\ref{sec:Equations} represent a set of differential equations that may in principle be solved to give the full time evolution of the disk, under the assumption that this evolution is well approximated as a sequence of steady-state solutions (see discussion of this assumption below). We consider two models: a single-component disk supported either by stars or by sBHs only (to build intuition), and then an evolving model in which a growing population of sBHs seeded by the progenitor stellar population changes the steady-state structure of the disk over time. 

Our procedure for solving the sBH-only disk model is similar to the one laid out in TQM05 and more recently developed in \citet{2024gangardt}. Beginning at the outer disk boundary with $r = R_{\rm{out}}$ and $\dot{M} = \dot{M}_{\rm{out}}$, and assuming a marginally stable disk (i.e. $\rho \propto \Omega^2$), we can solve for $\tau$ and $T_{\rm{eff}}$ as a function of $T$. The total pressure is uniquely determined by $\rho$ and $\dot{M}$, while the right hand side of Equation (\ref{eqn:Mdot}) yields $p_{\rm{tot}} = p_{\rm{tot}}(T)$, allowing us to solve for $T$. Note that at some radii multiple disk solutions exist. This was dealt with in TQM05 by assuming the lowest $T$ solution to be the most physically motivated. Near $R_{\rm{out}}$, we follow TQM05, but at interior annuli, we choose the solution with opacity $\kappa$ closest to that of the adjacent exterior radius. Our motivation for this choice is discussed in more detail in \S\ref{subsec:degenSolutions} below.

By assuming the disk is in LTE, the disk flux $\sigma_{b} T_{\rm{eff}}^4(T)$ can be used to calculate the total auxiliary heating necessary to support the disk via Equation (\ref{eqnTeff}), where we set $\dot{\Sigma}_{\star} = 0$ and solve for $Q_{\rm{sBH}}$. Equation $(\ref{eqn:sBH accretion})$ gives $\dot{m}_{\rm{sBH}} = \dot{m}_{\rm{sBH}}(\rho, H)$ from which we can calculate the total emitted radiation from an sBH. Using Equation (\ref{eqn:fesc}) we determine $f_{\rm{esc}}$ and solve Equation (\ref{eqn:QBHNew}) for $S_{\rm{sBH}}$ or the total number of sBHs per unit area. The final step is calculating the mass accretion from the sBHs as $\dot{\Sigma}_{\rm{sBH}} = S_{\rm{sBH}} \dot{m}_{\rm{sBH}}$, multiplying by the area of the outermost radial annulus, and subtracting from $\dot{M}_{\rm{out}}$ as in Equation (\ref{eqn:Mdot2}), to obtain the mass flux at the adjacent interior radius where we apply the same algorithm.

This procedure is applied at each radial annulus moving inward in the disk, until solutions maintaining a marginally stable disk require non-physical model parameters, namely, $Q_{\rm{sBH}} < 0$. At this point viscous heating alone provides sufficient support and we revert to a standard Shakura-Sunyaev disk model. Note that for the Shakura-Sunyaev disk, $T_{\rm{eff}}$ is given and $\rho$ is not. To solve for $\rho$ and $T$ in this case we balance Equations (\ref{eqn:ptot}) and (\ref{eqn:T}) using Scipy's \citep{2020virtanen} optimization routine `root.' Values for $\rho$ and $T$ from the adjacent exterior radii are used as input guesses. We use 500 equally-spaced logarithmic radial bins to model the disks. At this resolution we found that these steady-state disk models converge. 

Our second model assumes the AGN disk's evolution over time can be treated as a sequence of steady states that change in response to the accumulating population of remnant sBHs. We begin at $t=0$ with a TQM05 disk, fully supported by star formation. We construct this model in much the same was as the sBH only model, but solve for $\dot{\Sigma}_{\star}$ instead of $Q_{\rm{sBH}}$. Given $\dot{\Sigma}_{\star}$ and a time step $\Delta t$ and assuming $\dot{\Sigma}_{\star}$ is constant over $\Delta t$, we solve for the number of sBHs in each radial annulus according to Equations (\ref{eqn:ax}-\ref{eqn:sbh}). 

Having populated the disk with remnants, we solve for two solutions sets: (1) a marginally stable ($Q_{\rm{T}} = 1$) disk solution supported by both star formation and radiative emission from embedded sBHs and (2) a stable disk supported only by sBHs ($Q_{\rm{T}} \geq 1$ and $\dot{\Sigma}_\star = 0$). If the former requires negative pressure support from stars ($\dot{\Sigma}_\star < 0$) or the latter requires a volume density exceeding $\Omega^2/(\sqrt(2)\pi G$ the solutions are presumed to be not physical and the alternate solution set is chosen. In some cases both sets of solutions yield physical disk solutions. This can occur where degenerate, marginally stable starburst solutions exist or where stable solutions require unreasonably low $\Sigma$ such that the sBHs can support the disk at sub-Eddington accretion rates. Where both solutions exist, we choose the stable (sBH-only) solution set when the number density of sBHs exceeds the minimum number required to support the disk as found in our initial, sBH-only models. This minimum sBH number is referred to as $S_{\rm{sBH}_{\rm{min}}}$ in the schematic shown in Fig.~\ref{fig:implementation}.  

Once solutions are found across all radii, we jump to the next time step and calculate the distribution of sBHs. Note that after the first time step we must discretize Equation~(\ref{eqn:sbh}) according to: 
\begin{equation}
    S_{\rm{sBH}}(t_{n}) = \sum_{i = 0}^n {A_{\rm{x}}(t_i)}\int_{m_{\rm{TO}}(t_n - t_i)}^{m_{\rm{max}}}{m_{\star}^{-\delta} dm_{\star}}
\end{equation}
where 
\begin{equation}
    A_{\rm{x}}(t_i) = \frac{\dot{\Sigma}_{\star}(t_i)\Delta t_i}{ \int_{m_{\rm{min}}}^{m_{\rm{max}}} m_{\star}^{1-\delta}dm_{\star}}
\,.\end{equation}
This procedure is laid out explicitly in Fig. \ref{fig:implementation}. Note that in this figure we distinguish between marginally stable solutions including both star formation and sBH accretion and those supported by sBHs only with a subscript 1 and 2 respectively.

In our models we take the first time step to be $2.5$ Myr, approximately the delay between the start of star formation and the most massive stars evolving off the MS. The second and subsequent time steps are set to $\Delta t = 1$ Myr, although we find that time resolution has limited effect on our models so long as $\Delta t \leq 2.5$ Myr. We also enhance our radial resolution in these models, increasing $\Delta r$ in regions where enhanced star formation leads to steep drops in mass flux. We discuss these resolutions tests in more detail in \S\ref{subsec:resolution}. 

We opt to simplify our models significantly by assuming all sBHs have a mass of 10 ${\rm M}_\odot$. The initial {\it versus} final mass relation (IFMR), linking remnant sBHs to their respective progenitor stars, can vary significantly depending on assumptions made about stellar winds, supernovae, and chosen metallicity \citep{2015spera, 2017spera, raithel2018}. Assuming solar metallicity, \citet{raithel2018} predicts maximum sBH masses of 16 $\rm{M}_\odot$ while the models of \citet{2015spera} suggest sBH masses can reach $25 \rm{M}_\odot$ for high metallicity systems ($Z = 2 \times 10^{-2}$). Despite discrepancies, we expect  $m_{\rm{sBH}} \sim 10 {\rm M}_\odot$ to be a reasonable estimate for our purposes. \footnote{Although we expect our results to be robust against small changes in $m_{\rm{sBH}}$ it is useful to note that increasing the mass would have the dual effect of increasing the accretion rate and softening the spectrum of the individual sBH. The latter would lead to a slight decrease in $f_{\rm{esc}}$, slightly decreasing the effective radiative efficiency of embedded sBHs.}

Although we incorporate the feedback from sBHs in our disk models, we do not account for the accompanying sBH growth. If we were to account for sBH growth, the number ($S_{\rm{sBH}}$), accretion rate ($\dot{m}_{\rm{sbh}}$), and escape fraction ($f_{\rm{esc}}$) would be evaluated independently for sBHs of differing mass and Equations (\ref{SigmadotBH}) and (\ref{eqn:QBHNew}) would need to re-written as a sum of products. This added a layer of complication that slowed our computation but would not significantly enhance our results.
For sBH accretion rate set to the maximum, Eddington rate, an AGN lifetime of $t_{\rm{AGN}} = 10^8$ years represents $\sim 2.5$ e-folding times, or about an order of magnitude increase in mass (and accretion rate). While this enhances the importance of sBHs, 
we do not expect this to change our qualitative results, a point we discuss in more detail in \S{\ref{subsec:growthtime}} below.

\section{Steady State Disks: Single source models}
\label{sec:steadystate}

\begin{figure*}
\centering
 \includegraphics[width=6in]{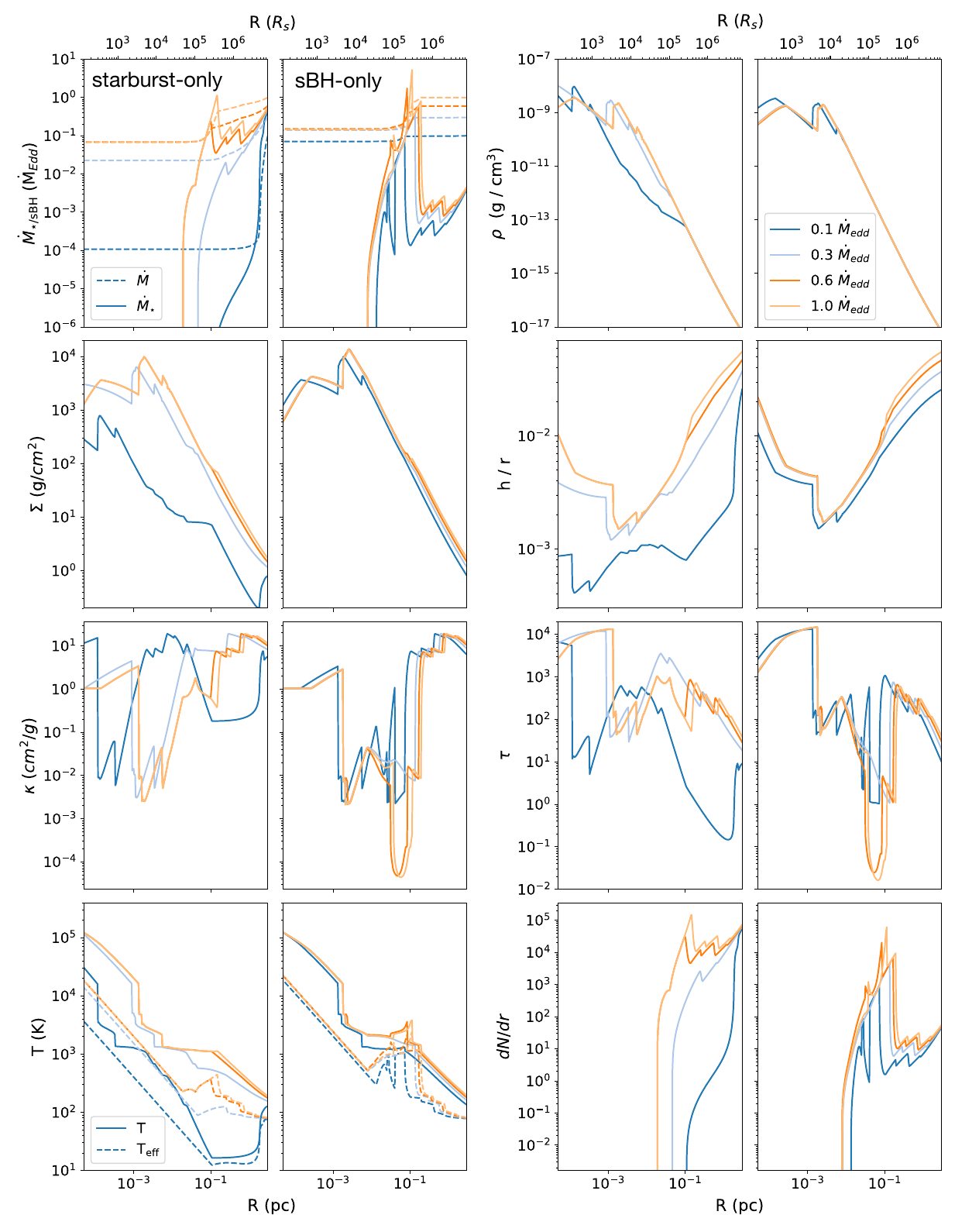}
 \caption{Disk parameters as a function of radius for models with $\text{M}_{\text{SMBH}} = 4\times10^6$ $\rm{M}_\odot$, $\sigma = 75$ km/s, and $R_{\text{out}} = 3$ pc. The outer boundary conditions $\dot{M}_{\rm{out}} = 0.1$, $0.3$, $0.6$, and $1$ $\dot{\rm{M}}_{\rm{Edd}}$ are differentiated by line color and correspond to dark blue, light blue, dark orange, and light orange respectively. The leftmost columns in each panel (i.e. the first and third column from left to right) show starburst-only solutions and the rightmost (second and fourth) columns show sBH-only disk solutions. From top to bottom the panels on the left show: mass flux through the disk ($\dot{M}$) and mass consumption by the auxiliary heat source, column density ($\Sigma$), opacity ($\kappa$), and temperature (solid) and effective temperature (dashed). From top to bottom the right panels show volume density ($\rho$)), dimensionless scale height ($h/r$), optical depth ($\tau$), and the number of stars/sBHs per radial annulus. The total number and mass of stars/sBHs is further documented in Table~\ref{tab:my_label}.}
    \label{fig:steadyState}
\end{figure*}
 
Here, we analyze steady-state AGN disks supported entirely by accretion onto sBHs. As noted in \S\ref{sec:Feedback}, we do not anticipate that sBH heating will quickly overcome stellar heating across the disk, nor do we expect the sBH population to grow to precisely the size necessary to maintain a marginally stable disk. However, by investigating the `sBH-only' case we can clarify expected disk profiles given a minimum number of sBHs, comparing directly with the TQM05 starburst-only models with equivalent boundary conditions. 

Our steady-state models are also analogous to the`pile-up' solutions found by \citet{gilbaum2022}. Note that the `pile up' solution assumes that the sBHs do not migrate and that mass flux is constant across the disk. The latter constraint is not assumed in the models presented here (we account for the reduction in the inflow rate as gas is consumed by BH accretion). Additionally, \citet{gilbaum2022} take the disk to be largely ionized such that electron scattering is the dominant source of X-ray opacity. Our models assume a neutral disk, opaque to X-ray flux emitted by sBHs when $\tau_{\rm{X}} > 1$. 

As described in \S\ref{subsec:XrayOpacityn}, $\tau_{\rm{X}}$ includes contributions from scattering and absorption by dust, neutral atoms, and photoionization. We refer to disk models including all of these X-ray/gas interactions as the `sBH-only' models. Here, we also present an alternative prescription for optical depth that includes only the effects of photoionization of hydrogen and singly ionized helium. These models are intended to address the possibility that the embedded BHs launch jets and evacuate the surrounding gas, creating a low-density "chimney" out of the disk~\citep{tagawa2022}.  As a result, the effective $\Sigma$ encountered by the BH radiation may be lower than that of the average background disk. Below $\Sigma\sim 10^{-2} \text{ g/cm}^2$, photoionization is the dominant contributer to $\sigma_{\rm{eff}}$ and other processes can be ignored. We refer to disks in which we used this amended $\tau_{\rm{X}}$ as `sBH-only-chimney' models and note that they have a larger $f_{\rm{esc}}$ than the `sBH-only' models. 

In the following subsections we describe these two classes of models, beginning with a description of relevant disk parameters as a function of distance from the central SMBH in \S{\ref{subsec:diskParams}}. We justify our assumption of a neutral gas disk in \S{\ref{subsec:heatMixing}} and address degenerate steady-state solutions in \S{\ref{subsec:degenSolutions}}. In \S{\ref{subsec:spectra}} we calculate the anticipated spectra for our steady state models and in \S{\ref{subsec:growthtime}} we estimate the timescale over which an AGN disk is fully supported by sBHs. 

\subsection{Star- vs sBH-driven Disk structures}
\label{subsec:diskParams}

\begin{table}
    \centering
    \begin{tabular}{|c|c|c|c|}
    \hline
        \quad & $\dot{M}_{\rm{out}} (\dot{\rm{M}}_{\rm{Edd}})$ & $N_{\rm{tot}}$ & $M_{\rm{tot}}$ ($\rm{M}_\odot$)\\
        \hline 
         stars-only & 0.1 & $4.5 \times 10^5$ & $1.0 \times 10^6$ \\
         \quad & 0.3 & $1.3 \times 10^6$ & $2.7 \times 10^6$ \\
         \quad & 0.6 & $2.4 \times 10^6$ & $5.2 \times 10^6$ \\
         \quad & 1 & $4.2 \times 10^6$ & $9.2 \times 10^6$ \\
        \hline
         sBHs-only & 0.1 & $1.2 \times 10^4$ & $1.2 \times 10^5$ \\
         chimney & 0.3 & $6.0 \times 10^4$ & $6.0 \times 10^5$ \\
         \quad & 0.6 & $1.7 \times 10^5$ & $1.7 \times 10^6$ \\
         \quad & 1 & $3.4 \times 10^5$ & $3.4 \times 10^6$ \\
         \hline
         sBHs-only & 0.1 & $1.2 \times 10^4$ & $1.2 \times 10^5$ \\
         \quad & 0.3 & $6.4 \times 10^4$ & $6.4 \times 10^5$ \\
         \quad & 0.6 & $1.8 \times 10^5$ & $1.8 \times 10^6$ \\
         \quad & 1 & $3.4 \times 10^5$ & $3.4 \times 10^6$ \\
         \hline
    \end{tabular}
    \caption{The total number and mass in stars or sBHs 
    accumulated over $10^8$ yrs
    in the steady-state models illustrated in Fig.~$\ref{fig:steadyState}$.}
    \label{tab:my_label}
\end{table}

In Fig.~\ref{fig:steadyState}, we show the radial profiles of different disk variables for starburst-only and sBH-only. Starburst-only profiles are on the leftmost (irst and third) columns while sBH-only profiles are in the rightmost (econd and fourth) columns. 

The mass-feeding rates span \(0.1 \, \dot{M}_{\text{Edd}}\) (dark blue) to \(1 \, \dot{M}_{\text{Edd}}\) (light orange). This range encompasses the critical value \(\dot{M}_{\rm{c}} = 0.2 \, \dot{M}_{\rm{Edd}}\) defined in Equation~(\ref{eqn:mc}), below and above which star-supported disk models exhibit fundamentally different behavior. By examining mass fluxes on either side of this bifurcation, we can understand how the behavior changes when disks with the same parameters are instead supported entirely by sBHs.

The upper left panels in Fig.~\ref{fig:steadyState} plot the mass flux through the disk (dashed) and mass consumption rates by stars (solid, left panel) and by sBHs (second and third-to-left panels). In the starburst-only case, there are two distinct disk solutions represented, differentiated by low (0.1 and 0.3 $\dot{\rm{M}}_{\text{Edd}}$) and high (0.6 and 1 $\dot{\rm{M}}_{\text{Edd}}$) $\dot{M}_{\text{out}}$. In the smaller $\dot{M}_{\rm{out}}$ cases the star formation rate peaks at the outer boundary leading to a steep drop in mass flux. Between 0.1 and 1 pc, $\dot{M}$ levels out, having dropped to $2 \times 10^{-2}$ and $1 \times 10^{-4}$ $\dot{\text{M}}_{\rm{Edd}}$ in the $\dot{M}_{\rm{out}} = 0.3$ $\dot{\rm{M}}_{\text{Edd}}$ and 0.1 $\dot{\rm{M}}_{\text{Edd}}$ cases, respectively. This behavior is characteristic of disks whose $\dot{{M}}_{\rm{out}}$ is approximately equal to or below the critical mass supply rate $\dot{M}_{\rm{c}}$, defined in Equation~(\ref{eqn:mc}). In contrast, the larger $\dot{{M}}_{\rm{out}}$ disk models maintain star formation down to $r=2\times10^{-2}$ pc, at which point star formation peaks and then plummets, and the mass supply rate drops to a constant value of $\simeq 7 \times 10^{-2}~\dot{\text{M}}_{\rm{Edd}}$. 

In the sBH-only models $\dot{M}_{\rm{out}}$ exceeds the critical mass supply rate in all of the cases considered, and there is no significant depletion in the outer disk. The change in mass flux across these disks is smaller than the starburst-only solutions in all cases, due to the higher radiative efficiency of sBHs relative to stars. Despite distinct $\tau_{\rm{x}}$ and $f_{\rm{esc}}$, the sBH-only and sBH-only-chimney models have very similar radial profiles. This similarity is reflected in the intrinsic disk spectra shown in Fig. \ref{fig:spectra} and the similar numbers of sBHs necessary to support the disks (Table \ref{tab:my_label}). In both sets of models, the mass flux through the disk drops to $\simeq 0.15 ~\dot{\rm{M}}_{\rm{Edd}}$ in the three highest $\dot{{M}}_{\text{out}}$ cases, while in the lowest $\dot{{M}}_{\text{out}}$ case the mass flux reaching the SMBH is $\simeq 7 \times 10^{-2} ~\dot{\rm{M}}_{\rm{Edd}}$. Because the disk is not significantly depleted in either of the sBH cases, the volume density ($\rho$), column density ($\Sigma$), and scale height ($h$) all remain comparable to the higher $\dot{M}_{\rm{out}}$, starburst-only cases.

In the bottom right panels of Fig.~\ref{fig:steadyState}, we plot the number of stars and sBHs per radial annulus. The number of stars is calculated assuming a constant $\dot{\Sigma}_\star$ over an AGN lifetime of $t_{\rm{AGN}} = 10^8$~yr. We give the total mass and number of stars and sBHs in Table~\ref{tab:my_label}. Note that $M_{\rm{tot}}$ in the sBH-only and sBH-only-chimney models never exceeds the mass of the SMBH, justifying the continued use of a Keplerian potential. This is not true for the higher $\dot{M}_{\rm{out}}$, starburst-only cases. 

There are also slightly more sBHs in the $\dot{M}_{\rm{out}} = 0.3$ and $0.6 ~\dot{\rm{M}}_{\rm{Edd}}$ sBH-only models than in their sBH-only-chimney counterparts. This result is somewhat counter-intuitive given that in the chimney scenario more sBHs are required to produce the same amount of heating. Exterior to the opacity gap ($r \gtrsim  0.1 $ pc) the number of sBHs in the chimney models is greater than in the sBH-only models, as expected. The result is a slight increase in mass flux through the opacity gap in the sBH-only models relative to the chimney models. This slight increase in mass flux in the disk interior requires a larger number of sBHs in the sBH-only case to stabilize the disk, thereby explaining the slight enhancement in number in the sBH-only case.

\subsection{Heat mixing in a multi-phase disk}
\label{subsec:heatMixing}

The homogeneity of the disk fluid depends on the number of sBHs embedded in it relative to the volume of gas the sBHs can be expected to ionize. This latter value can be approximated by the volume within the Str\"omgren radius,
\begin{equation}
\begin{split}
R_{\rm{Str}} =& \left( \frac{\int_{h_{\rm p} \nu_1}^{\infty} \frac{L_{\rm{sBH}_\nu}}{h_{\rm p} \nu} d \nu}{\frac{4 \pi}{3} n_{\rm{H}}^2 \alpha_{\rm{B}}}\right)^{1/3} \\
&\simeq 1.9 \times 10^{-4} ~\text{pc} \left(\frac{\rho}{10^{-16} \text{ g/cm}^3}\right)^{-\frac{2}{3}}\,.
\label{eqn:Rstr}
\end{split}
\end{equation}  
Here $\alpha_{\rm B} = 2.59 \times 10^{-13}$ is the Case B recombination cross-section for HI \citep{2006osterbrock}, $n_{\rm{H}} = \rho/(X m_{\rm{p}})$ is the number density of Hydrogen, and $X = 0.75$ is the mass fraction of Hydrogen for primordial gas. We parameterize $R_{\rm{Str}}$ in terms of $\rho$ by assuming Eddington-capped accretion.  

\begin{figure}
\centering
 \includegraphics[width=3.3in]{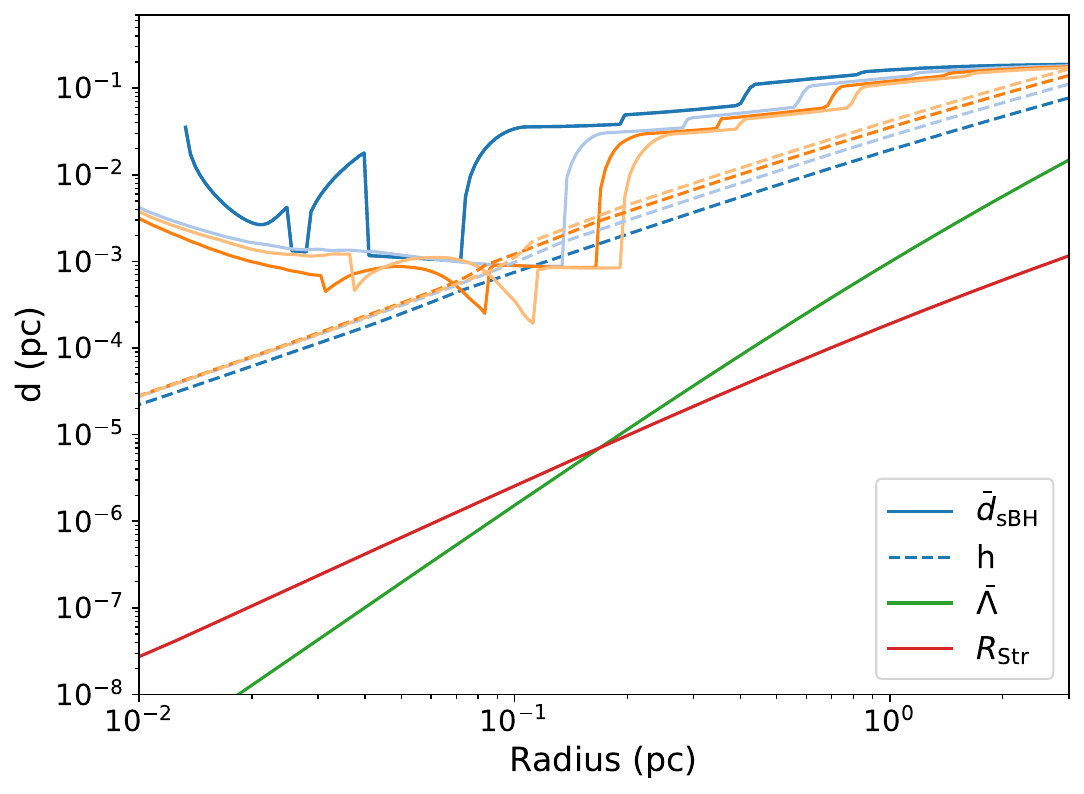}
 \caption{The scale height $h$ (dashed) and average distance between sBHs $\bar{d}_{\rm{sBH}}$ for sBH-only models. As in Fig.~\ref{fig:steadyState}, boundary conditions are differentiated by line colors: dark blue, light blue, dark orange, light orange. The Str\"omgren radius $R_{\rm{Str}}$ is plotted in red and the average mean free path $\bar{\Lambda}_{\rm{photo}}$ is plotted in green. $R_{\rm{Str}}$ and $\bar{\Lambda}_{\rm{photo}}$ never exceed  $\bar{d}_{\rm{sBH}}$, justifying our assumption of a neutral disk.}
    \label{fig:stromgren}
\end{figure}

The Str\"omgren radius approximates the size of the ionized bubble around a single sBH, assuming the ionization profile is a step function 
(i.e. the gas is completely ionized interior to $R_{\rm{Str}}$ and completely neutral beyond). If the mean free path of the average ionizing photon is long, the ionization profile flattens and the boundary of the ionized bubble is blurred~\citep{KramerHaiman2008}. The ionizing mean free path $\Lambda_{\rm{ion}}$, is calculated in much the same way as $\Lambda_{\rm{X}}$ in \S\ref{subsec:XrayOpacityn}, but does not include absorption by dust. Dust absorption reduces the ionizing emission but does not contribute to the ionization fraction. That is,
\begin{equation}
\begin{split}
    \Lambda_{\rm{ion}} = &\left(n_{\rm{H}}\sigma_{\rm{HI}} + n_{\rm{He}}\sigma_{\rm{HeI}}\right)^{-1} \\
    & \times \left[ \sqrt{1 + \frac{n_{\rm{H}}(1.5\sigma_{\rm{C}} + \sigma_{\rm{DS}})}{(n_{\rm{H}}\sigma_{\rm{HI}} + n_{\rm{He}}\sigma_{\rm{HeI}})}} \right]^{-1}\,.
\end{split}
\end{equation}
The ionizing photon-weighted, average mean free path is then calculated according to, 
\begin{equation}
\begin{split}
    \bar{\Lambda}_{\rm{ion}} \equiv &  \frac{\int_{h_{\rm p} \nu_1}^{\infty}  \frac{L_{\rm{sBH}_\nu}}{h_{\rm p} \nu} \exp{(-\tau_{\rm{DA}})} \Lambda_{\rm{ion}} d \nu}{ \int_{h_{\rm p} \nu_1}^{\infty}  \frac{L_{\rm{sBH}_\nu}}{h_{\rm p} \nu} \exp{(-\tau_{\rm{DA}})}d\nu} \\ 
    & \simeq 9.9 \times 10^{-4}~\text{pc}\left(\frac{\rho}{10^{-16}\text{ g/cm}^3}\right)^{-1}\,,
\label{eqn:avgMFP}
\end{split}
\end{equation}
where we have assumed the ionizing flux is attenuated due to dust absorption according to $\tau_{\rm{DA}} = n_{\rm{H}} \sigma_{\rm{DA}}h$, 
and sBH accretion is Eddington capped. We can assess the importance of this blurring by comparing 
$\bar{\Lambda}_{\rm{ion}}$ 
with the Str\"omgren radius. Setting $R_{\rm{Str}} = \bar{\Lambda}_{\rm{ion}}$ we can solve for $\rho \simeq 1.4\times 10^{-14}~{\rm g~cm^{-3}}$. For a marginally stable disk, this density translates to a distance of $\simeq 0.17$ pc. We plot Equations (\ref{eqn:Rstr}) and (\ref{eqn:avgMFP}) as a function of distance in Fig.~\ref{fig:stromgren}. For $r< 0.17$ pc, $R_{\rm{Str}} > \bar{\Lambda}_{\rm{photo}}$ suggesting that the sBHs carve out distinct, ionized bubbles in the neutral disk fluid. Beyond $0.17$ pc, we expect the gas around embedded sBHs to have a broadened ionization profile. 

Fig.~\ref{fig:stromgren} also shows the average distance between sBHs, $\bar{d}_{\rm{sBH}} = S_{\rm{sBH}}^{-1/2}$, and the scale height $h$ in all four sBH-only models. We can conclude that ionization bubbles do not typically overlap, because $\bar{\Lambda}_{\rm{photo}} < \bar{d}_{\rm{sBH}}$, although these lengths become comparable at large radii. This finding is consistent with our discussion in \S\ref{subsec:XrayOpacityn}, in which we found larger escape fractions in the outer disk, where $\bar{\Lambda}_{\rm{photo}}$ approaches $h$. Most importantly, $R_{\rm{Str}} \ll \bar{d}_{\rm{sBH}}$ across all sBH-only models and radii, indicating that the disk may be treated as neutral and the effects of electron scattering can be neglected. Moreover, we expect photons produced by recombination will fall primarily in the IR and optical bands, where their mean free path is longer, and they can be assumed to uniformly heat the disk, although the details of these processes need to be addressed in future work.

\subsection{Degenerate solutions}
\label{subsec:degenSolutions}

\begin{figure}
\centering
 \includegraphics[width=3.4in]{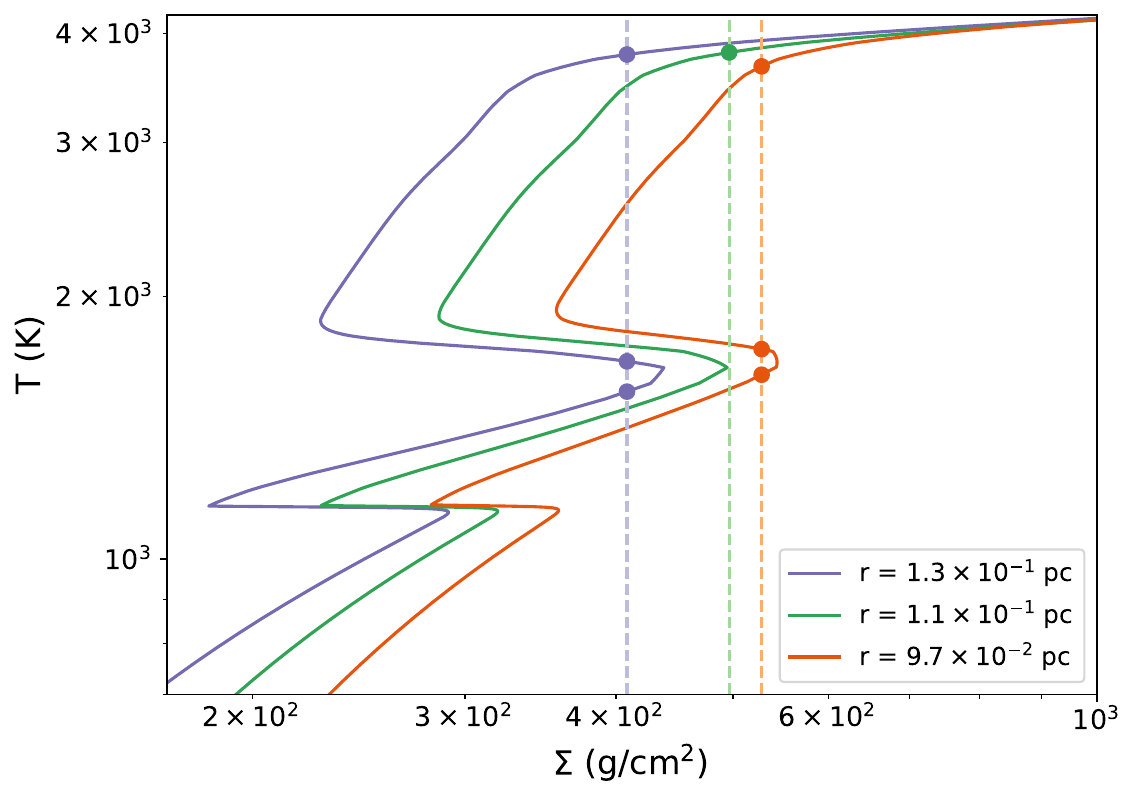}
 \caption{Thermal equilibrium curves in the ($\Sigma$,$T$) plane for the sBH-only models where $\dot{M}_ {\rm{out}} = 1.0$ $\dot{\rm{M}}_{\rm{Edd}}$. Solid curves show equilibria at nearby radii: $1.3 \times 10^{-1}$ pc (purple),$1.1 \times 10^{-1}$ pc (green), and $9.7 \times 10^{-2}$ pc (orange). Vertical dashed lines indicate the
 $\Sigma = \Sigma(r, \dot{M})$ derived from the steady-state model. 
 Finally, dots marks the solution or set of solutions at each radius. Steady-state solutions exhibit discontinuous jumps across varying $r$, with low temperature solutions ($\sim 1.8 \times 10^3$ K) available at $r = 1.3 \times 10^{-1}$ pc and $r = 9.7 \times 10^{-2}$ pc but not $1.1 \times 10^{-1}$ pc.}
    \label{fig:T_Sigma}
\end{figure}

In the starburst-only, sBH-only, and sBH-only-chimney disk models, there are ranges of $\dot{M}$ and $r$ across which multiple steady-state solutions exist. Typically, these solutions can be categorized into three branches, a low-temperature ($\sim 1000 $~K) `cold' solution, a high-temperature `hot' solution ($\sim 2000 - 3000$~K) and an intermediate, thermally unstable solution. This multibranch behavior is classically associated with outbursts in dwarf novae, manifesting at around $7000$~K due to a steep rise in opacity linked to $\text{H}^-$ scattering (e.g. \citealt{2001lasota}). In contrast, we find degenerate solutions in the cooler, outer regions of AGN -- straddling the opacity `bump' from molecular absorption by water vapor at $\sim 2000$ K.

The existence of multiple solutions in the starburst-only disk models has been noted across previous works. TQM05 approached the degeneracy by prioritizing the lowest temperature solution in all cases. They argue that both the intermediate- and high-temperature solutions are likely to be unstable and can therefore be discarded, using formal stability arguments in the intermediate solution case and qualitative physical arguments in the high temperature case. In the starburst-only disk models discussed here we follow the TQM05 approach, choosing the `cold' solutions when degeneracies arise.

Following the same approach in the sBH-only and sBH-only-chimney cases, however, does not always yield physical results. At $\lesssim 0.1$ pc in the two highest $\dot{M}_{\rm{out}}$ cases, the high mass flux forces the disk from the degenerate solution regime onto the 'hot,' optically thin branch. Because $T_{\rm{eff}} \simeq T \tau^{-1/4}$, $T_{\rm{eff}}$ increases dramatically along with accretion onto sBHs ($\dot{\Sigma}_{\rm{sBH}}$) which scales with $T_{\rm{eff}}^4$. The steep decrease in $\dot{M}$ that results leads to reappearance of the low-temperature solution branch at interior radii.

This situation is illustrated in Fig.~\ref{fig:T_Sigma}, where three thermal equilibrium curves at $r = 1.3 \times 10^{-1}$ pc (purple), $r =1.1 \times 10^{-1}$ pc (green), and $r = 9.7 \times 10^{-2}$ pc (orange) are shown in the ($\Sigma$,$T$) plane. The nontrivial shapes of the curves in Fig.~\ref{fig:T_Sigma} follow from the temperature-dependence of the opacity. Points marks the steady-state solution(s) along these curves. Note that as we move from the outermost to the innermost radius,
we begin with three possible solutions ($1.3 \times 10^{-1}$ pc), drop to one solution ($r = 1.1 \times 10^{-1}$ pc), and then jump back to three solutions again ($9.7 \times 10^{-2}$ pc). 
If we were to choose the lowest temperature solution in all cases, our modeled disk solutions would oscillate between the cold and hot solution branches. Physically, we do not expect such abrupt jumps in temperature and opacity. Although not taken into account formally here, disks experience some radial heat transport through radiation and turbulence, very likely smoothing such extreme variations in opacity and temperature. In an effort to avoid this highly oscillatory and unphysical radial profile, we choose the solution whose opacity is closest to that found in the adjacent exterior radius. Thus, in the high $\dot{{M}}_{\text{out}}$ cases, once the jump to the hot branch of the solutions is made, a high temperature and low opacity is maintained in the disk interior. 

In the lowest $\dot{M}_{\rm{out}}$ case the same process leads to jumps between `cold' and `hot' solutions, although shifted slightly towards the SMBH -- at a distance of $r \lesssim 5 \times10^{-2}$ pc. When $\dot{M}_{\rm{out}} = 0.3 ~\dot{M}_{\rm{Edd}}$ the disk profiles are the same regardless of which solution-picking algorithm we use. 


\subsection{Emerging spectra}
\label{subsec:spectra}

\begin{figure}
\centering
 \includegraphics[width=3.1in]{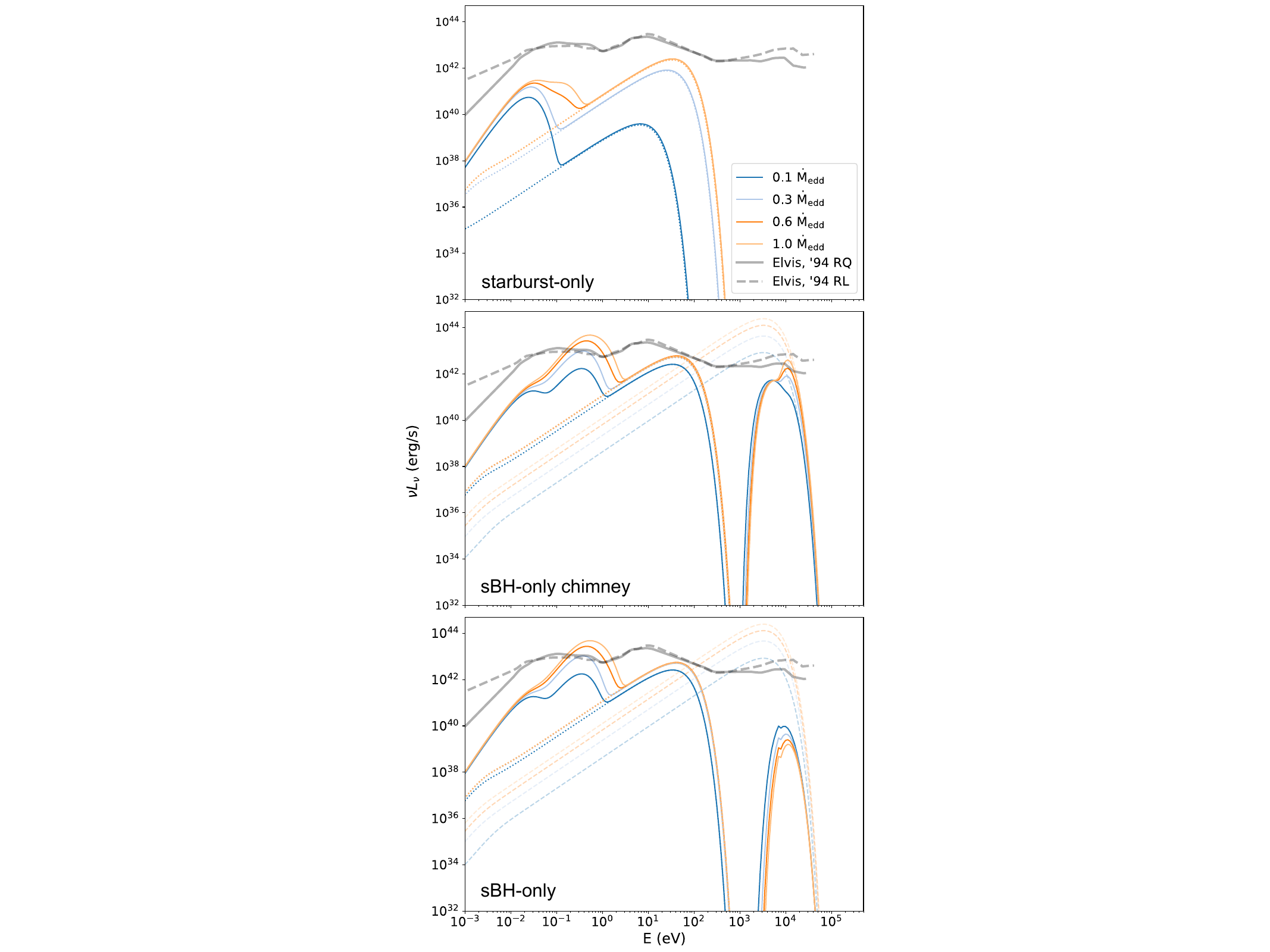}
 \caption{Spectral energy distributions (SEDs) emerging in the starburst-only (top panel), sBH-only-chimney (middle panel), and sBH-only (bottom panel) disk solutions presented in Fig.~\ref{fig:steadyState}. Solid curves show the composite thermal spectrum emitted by the AGN disk. In the sBH-only and sBH-only-chimney models the solid spectrum also includes the escaping SED of the embedded sBH population -- visible as a high energy bump peaking at $\sim 10$ keV. In all three panels, the dotted curves show the SED of a standard Shakura-Sunyaev $\alpha$-disk, with an $\dot{M}$ matching the inner mass flux of each model and $\alpha = 0.3$. The dashed lines in the bottom two panels illustrate the total contributed sBH flux. We discuss these modeled spectra in terms of several component features, including: a mid-IR bump ($E \lesssim 0.1$ eV), a near-IR bump ($0.1~\rm{eV} \lesssim E \lesssim 1$ eV), a UV-bump ($1~\rm{eV} \lesssim E \lesssim 20$ eV), and an X-ray bump that peaks at $\sim 10$ keV. The gray lines show the averaged observed emission from radio quiet (solid) and radio loud (dashed) AGN taken from \citet{1994elvis}, normalized by a factor of $5\times10^{-3}$, such that the integrated strength of UV emission is $L_{\rm{UV}} = \rm{L}_{\rm{Edd}} \simeq 5.8 \times 10^{44}$ erg/s.}
    \label{fig:spectra}
\end{figure}

AGN are observed to produce significant flux, spanning nearly nine decades in frequency, from radio to X-ray \citep{1994elvis}. Emission across such a broad wave-band must arise from distinct regions and physical mechanisms in the disk. Several continuum features are seen consistently across observed AGN spectra, and have been used to constrain sources of emission, including: an increase in emission in the optical-ultraviolet known as the `UV-bump,'  flux exceeding classic $\alpha$-disk models in the radio, IR and X-ray bands, and an 
inflection point at $\sim$1 eV or $\sim 1 ~\mu$m . 

In Fig.~\ref{fig:spectra} we show the spectra of the starburst, sBH-only-chimney, and sBH-only disk models (top, middle, and bottom panels respectively), computed assuming that every annulus in the disk radiates as a black body. In the preceding subsections, We examine the UV, IR and X-ray emission predicted by our models and compare our results with the mean spectral energy distribution (SED) for radio quiet (RQ) and radio loud (RL) quasars from \citet{1994elvis}. These reference spectra are plotted in gray, and scaled by a factor of $5 \times 10^{-3}$. This scaling corresponds approximately to the ratio of the integrated UV luminosity of the \citet{1994elvis} data to the Eddington luminosity of a $4\times10^6 ~\rm{M}_{\odot}$ BH. This provides a useful benchmark for comparison against our relatively low-mass SMBH models.


\subsubsection{The UV-bump}
\label{subsubsec:UV}

The UV-bump is conventionally ascribed to thermal emission from the viscously heated disk interior \citep{1983malkan, 1987czerny} with energies $\gtrsim 1$ eV. The relationship between the classic thin disk spectrum and the UV-bump present in our model spectra is illustrated explicitly by plotting the Shakura-Sunyaev disk spectra (dotted lines) with $\alpha = 0.3$ and mass flux equal to $\dot{M}(R_{\rm{in}})$ of our corresponding model spectra. Critically, the Shakura-Sunyaev spectra are indistinguishable from their sBH/starburst supported counterparts in the UV-bump domain, suggesting that this feature is parameterized entirely by the mass flux ($\dot{M}$) close to the SMBH and the mass of the SMBH.

Note that the UV-bump predicted by our models is bluer than the bump present in the \citet{1994elvis} composite spectra. This discrepancy may be explained by recalling that Shakura-Sunyaev disks have 
a maximum temperature of $k_{\rm{B}}T \sim 10 (\dot{m}/10^8 \rm{M}_\odot)^{1/4}$ eV where $\dot{m} = \dot{M}/\dot{\rm{M}}_{\rm{Edd}}$ \citep{shakura1973}. The relatively low mass of our adopted SMBH, 
which is a factor of $\sim 100$ times lower than the typically $10^8-10^9~{\rm \rm{M}_\odot}$ SMBHs powering the bright quasars in the \citet{1994elvis} sample, can explain the apparent shift of our UV-bump to a factor of 2-3 higher energies.

The spectra shown in Fig. \ref{fig:spectra} are also faint relative to most observed AGN. With bolometric luminosities of $\sim 10^{-2} ~\rm{L}_{\rm{Edd}}$ or $5\times10^{44}$ erg/s, these models are significantly less luminous than typical bright AGNs, which emit closer to their Eddington limit with luminosities of $\sim 10^{46}$ erg/s. The relatively low luminosities we see in Fig. \ref{fig:spectra} are not uncommon for disks surrounding low mass SMBHs \citep{2007greene}, but suggest that these results cannot be extrapolated to higher-mass systems whose accretion rates commonly reach or exceed the Eddington limit. Interestingly, AGN emitting at a lower fraction of their Eddington rate have distinct spectra characterized by the absence of a 'big blue bump' and instead exhibit a broad excess in the mid-IR (e.g. \citealt{2008ho} and references therein); these are spectral features that are somewhat better aligned with our modeled spectra, but also indicative of their limited scope. The dependence of these models on SMBH mass is discussed further in \S\ref{sec:HighMass}.


\subsubsection{IR emission}
\label{subsubsec:IR}

The mass supply rate and the spectrum have a much less direct relationship in the IR regime, where starburst/sBH heating exceeds viscous heating and sets $T_{\rm{eff}}$. The result is IR emission well in excess of a standard steady Shakura-Sunyaev $\alpha$-disk. Our starburst-only models align with results from TQM05. Low mass supply rates ($\dot{M}_{\rm{out}} < \dot{M}_{\rm{crit}}$) result in depleted inner disk mass flux and a correspondingly small, red-shifted UV-bump. In the $\dot{M}_{\rm{out}} = 0.1 ~\dot{{M}}_{\rm{Edd}}$ case, flux from the outer, star forming region actually dominates thermal emission -- contrary to observations. Higher $\dot{M}_{\rm{out}}$ models have a broader and subdominant IR-bump. The broadening of the IR peak is the result of additional star formation and increased $T_{\rm{eff}}$ in the opacity gap at $\sim$ 0.1 pc.

At energies between $10^{-3}$ eV and $1$ eV, observed emission also diverge from the classic $\alpha$-disk, with an apparent emission excess in the IR. TQM05 concluded that this emission was broadly consistent with their modeled results. This excess is, however, commonly attributed to the reprocessing of inner disk flux by dust in a circumnuclear torus \citep{1993antonucci}. Still, as of yet, there is no universally accepted model template for torus emission. Additionally, factors such as contamination from host galaxy IR light and complex torus models with numerous degrees of freedom have led to a wide range of results \citep{2015netzer}. Several models have been proposed in which the source of IR emission is a warped or flared disk \citep{1989sanders, 2018baskin, 2023landt} and imaging by the GRAVITY collaboration suggests that, in the case of NGC 1068 the near-infrared emission is local to a thin disk \citep{20200pfuhl}.

Additional constraints on the near-IR (NIR) emission come from reverberation mapping studies in AGN. These studies show that the NIR bump typically lags the optival/UV continuum by tens to hundreds of days \citep{2006suganuma, 2014koshida}. These results are suggestive of dust heated and reprocessed by a circumnuclear torus as opposed to local emission by stars/sBHs in the disk. However, reverberation mapping campaigns have limitations in accurately separating the accretion disk contribution from toroidal dust emission. This limitation has been recently emphasized in the spectroscopic reverberation studies by \citet{2019landt} and \citet{2023landt}, in which a `constant red component' is observed in the NIR spectrum of NGC 5548 and Mrk 876. This component is separate from the variable emission attributed to hot dust and represents an enhanced emission in the NIR over the standard $\alpha$-disk spectrum. This behavior is consistent with auxiliary heating from a disk-embedded population.

In our sBH-only and sBH-only-chimney models, emission in the IR is further enhanced, because the sBHs allow a higher mass inflow rate into the inner regions of the disk. The result is a larger surface density interior to $0.1$ pc, requiring a higher heating rate to stabilize this region. The resulting IR bump is shifted to higher energies -- dominated by the near, not the mid, IR. Higher $\dot{M}_{\rm{out}}$ leads to further enhancement of the NIR relative to the rest of the spectrum, making it the dominant spectral feature in the $\dot{M}_{\rm{out}} = 0.3, 0.6$, and $1.0 ~\dot{\rm{M}}_{\rm{Edd}}$ sBH models. As illustrated in the \citet{1994elvis} composite spectra, such strong NIR emission is not commonly observed in AGN -- indicating that sBH-only supported disks are rare. This constraint is discussed in more detail in \S\ref{subsec:Sequence Spectra}. 


\subsubsection{X-ray emission}
\label{subsubsec:X-ray}

Hard X-ray emission ($ > 2$ keV) in AGN has been widely attributed to inverse Compton scattering of disk photons in a hot corona sandwiching the AGN \citep{1976shapiro, 1980sunyaev, 1991Haardt}. The emission is characterized by a power-law shape between 2 and 10 keV. These X-ray photons may also reflect off the nearby disk, creating the so-called Compton hump, an excess of hard X-rays over the power-law continuum, peaking between $\sim$20 - 30 keV \citep{2015netzer}. The observed spectrum is therefore the sum of primary continuum and reprocessed X-rays.
 
\begin{figure}
\centering
 \includegraphics[width=3.1in]{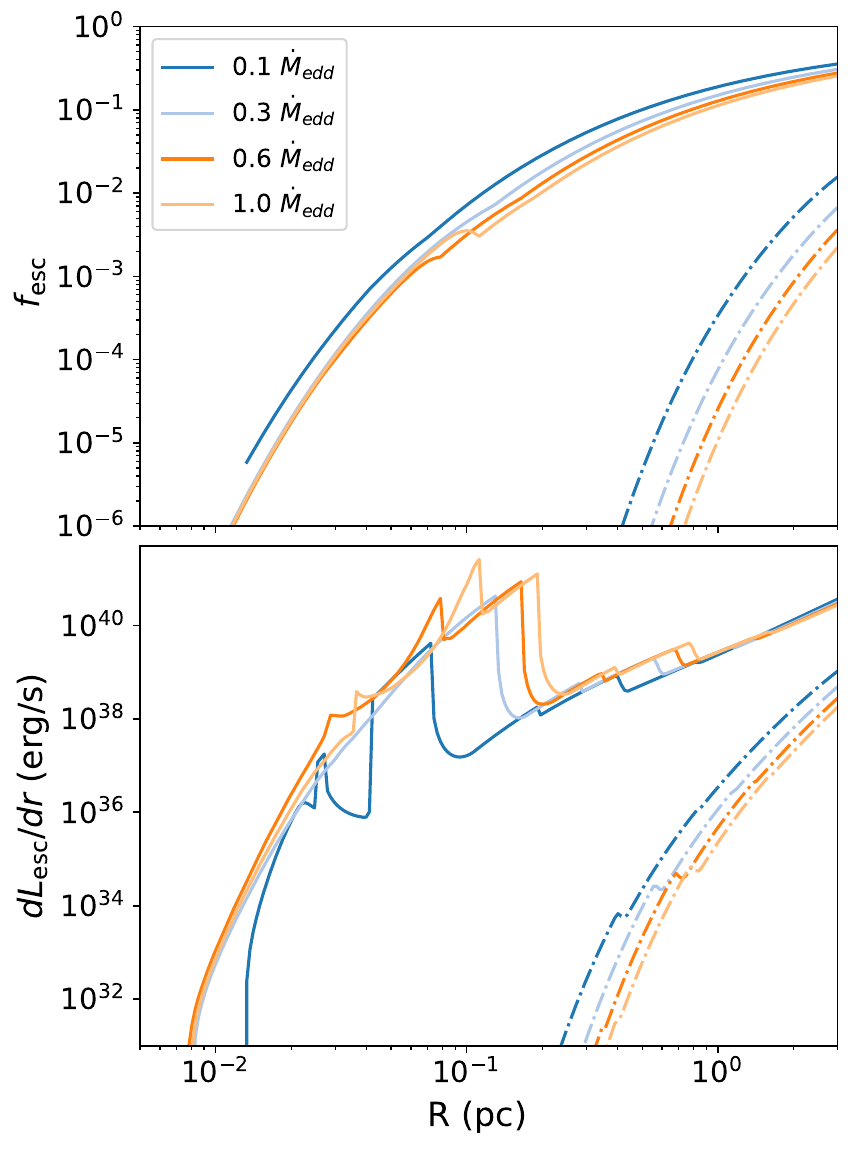}
 \caption{Escape fraction $f_{\rm{esc}}$ (top panel) and escape luminosity per log-radial bin $dL_{\rm{esc}}/ dr$ (bottom panel) as a function of distance from the central SMBH for the sBH-only-chimney models (solid lines) and sBH-only models (dot-dashed lines). Across all disk models $f_{\rm{esc}}$ is largest in the outer disk. However, the large sBH population in the opacity gap region combined with a slightly higher $f_{\rm{esc}}$ means that a $dL_{\rm{esc}}/dr$ is still significant in the chimney models at distances between $2\times 10^{-2} - 2\times 10^{-1}$ pc.}
    \label{fig:f_esc_L_esc}
\end{figure} 

In the bottom two panels of Fig.~\ref{fig:spectra} there is a spectral bump at $\sim 10$ keV due to emission from embedded sBHs escaping the disk. From lowest to highest $\dot{M}_{\rm{out}}$, the ratio of the escaping sBH flux to the bolometric luminosity in the sBH-only-chimney models is $6.9 \times 10^{-2}$,  $3.4 \times 10^{-2}$, $2.6 \times 10^{-2}$, and $2.8 \times 10^{-2}$. In the sBH-only models the ratios are $9.3 \times 10^{-4}$, $1.2 \times 10^{-4}$, $3.1 \times 10^{-5}$, and $1.1 \times 10^{-5}$. The significant contribution of X-ray flux to bolometric luminosity in the chimney case is consistent with the lower optical depth and enhanced $f_{\rm{esc}}$ in these models. Aside from this additional X-ray flux, the intrinsic disk flux of the sBH-only and sBH-only-chimney models is approximately the same. 

The escaping sBH flux peaks just beyond the low-energy edge of the Compton hump ($\sim$10 keV). The apparent absence of an observed emission excess at this energy in typical AGN spectra, beyond what can be attributed to coronal emission or Compton upscattering, suggests a constraint on the number of sBHs expected in AGN. Moreover, the hard X-ray emission from sBHs should differ from intrinsic AGN X-ray emission in several important ways. Notably, Compton hump emission is expected to be polarized \citep{2023Podgorny}, experience reverberation lags coincident with coronal variability \citep{2014uttley, 2015kara, 2021zoghbi}, and may be constrained to the inner disk \citep{1989fabian, 2004miniutti} -- none of which apply to the flux emitted from the sBHs modeled here.

In Fig.~\ref{fig:f_esc_L_esc} we plot $f_{\rm{esc}}$ (top panel) and $L_{\rm{esc}}$ (bottom panel) as a function of disk radius for the sBH-only-chimney (solid lines) and sBH-only (dot dashed lines) models. Note that in the sBH-only models the escaping X-ray emission is limited to the outer disk, beyond $0.1$ pc and the opacity gap, while in the sBH-chimney models, $L_{\rm{esc}}$ persists out to $0.01$ pc. Note that the energy of escaped emission is primarily a function of $\Sigma$, which ranges from $10^{-1} ~\rm{to} ~10^{1}~\rm{g/cm}^2$ at these radial distances. X-rays escape when $\tau_{\rm{X}} = 1$ or $\sigma_{\rm{X}} \sim 10^{-23}$ to $10^{-25} ~\rm{cm}^2$. From Fig. \ref{fig:xsection} we see that these cross sections correspond to energies straddling 10 keV -- confirmed by the location of the X-ray peak in Fig. \ref{fig:spectra}.Both because it is constrained to the outer disk and is sourced from upwards of 100 individual sBHs, the escaped X-ray flux should, on average, exhibit minimal variability over long timescales. In contrast, hard X-ray emission observed in AGN typically varies on very short timescales -- years \citep{2014Soldi} or even hours \citep{2012Ponti} -- further suggesting the X-ray emitting region is compact (e.g. \citealt{2013DeMarco, 2020ursini}). Additionally, the fractional root mean square variability amplitude can be as a high as 50\% \citep{2012caballero}. To ensure sBH emission does not dominate AGN emission in the 5-10 keV band or impact variability significantly at those energies, we estimate that sBHs can contribute only $\sim$1 - 10\% of observed X-ray emission. Assuming intrinsic AGN X-ray luminosity is $\sim$10\% of the bolometric luminosity, we conservatively estimate a normalized sBH contribution limit of $\sim 10^{-3} ~L_{\rm{bol}}$.

This limit is violated by our chimney models, which predict an excess X-ray emission exceeding $10^{-2} ~L_{\rm{bol}}$. The absence of such detectable excess in observations strongly suggests that the chimney models cannot accurately describe the behavior of embedded sBHs in AGN disks. Note too that the relative strength of the X-ray peak in the low $\dot{M}_{\rm{out}}$ sBH-only cases indicated that constraining our models may be more feasible by examining the spectra of low-luminosity AGN.The prediction of an X-ray emission excess in AGN spectra from sBHs suggests a new observational constraint on the number of sBHs embedded in the disk and further study of X-ray flux from AGN is likely to place tight upper limits on the number of sBHs present in the outer radii of AGN disks. A more thorough numerical analysis, incorporating additional physics such as coronal emission and other deviations from thermal thin-disk spectra in the emission spectra of sBHs, is left to future work.


\subsection{sBH population growth timescale}
\label{subsec:growthtime}

\begin{figure*}
\centering
 \includegraphics[width=7in]{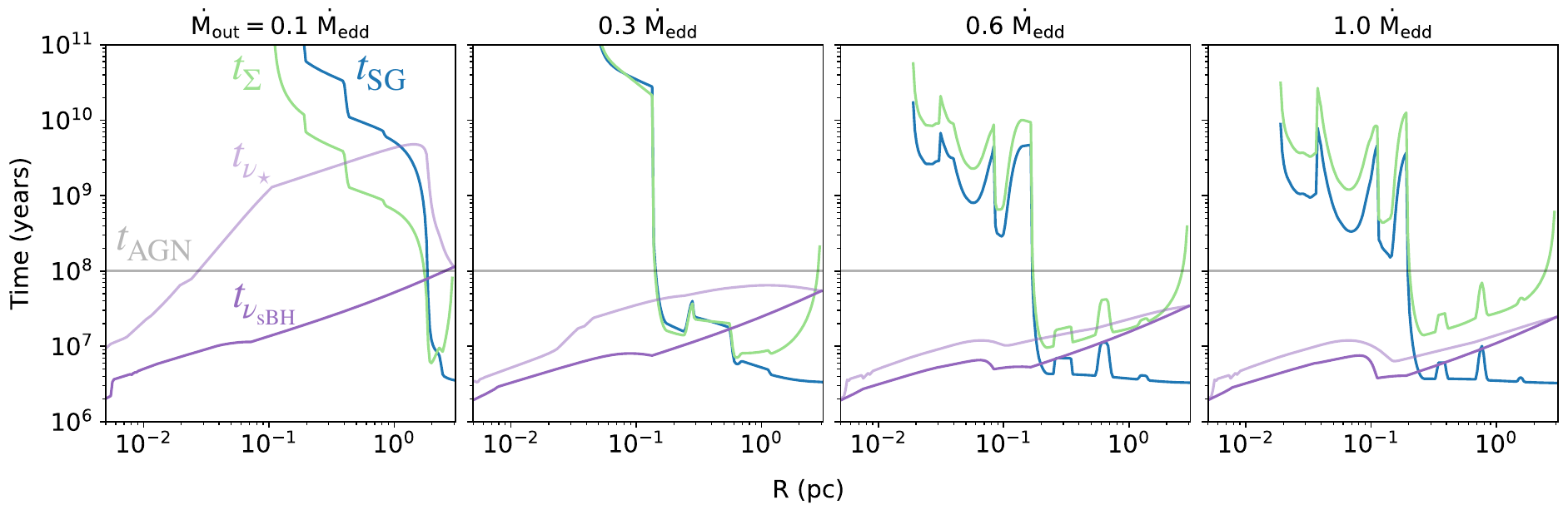}
 \caption{Anticipated timescales for sBH population growth $t_{\rm{SG}}$ (blue) and surface density change $t_{\Sigma}$ assuming that the disk is initially in a starburst-only state and evolves toward the sBH-only state. The gray line marks the expected lifetime of the disk $t_{\rm{AGN}} = 10^8$ years. Light and dark purple lines indicate the viscous timescales for the starburst-only and sBH-only disk respectively. From left to right, panels show these timescales for the $\dot{M}_{\rm{out}} = 0.1 ~\dot{\rm{M}}_{\rm{Edd}}$, $0.3 ~\dot{\rm{M}}_{\rm{Edd}}$, $0.6 ~\dot{\rm{M}}_{\rm{Edd}}$, and $1.0 ~\dot{\rm{M}}_{\rm{Edd}}$. These timescales provide a useful point of reference for the evolving models described in \S\ref{sec:evolution}.}
    \label{fig:Timescales}
\end{figure*}

The steady-state models developed in this section treat stars and sBHs independently, but our goal is to understand how the rate of star formation responds to the growth of the sBH remnant population. From this perspective, the starburst-only model can be viewed as the `initial state' of the AGN disk while the sBH-only disk approximates a hypothetical `final state.' The time required to evolve from the initial to the final state is calculated by assuming a constant star formation rate, according to
\begin{equation}
\begin{split}
t_{\rm{SG}} =& \frac{S_{\rm{sBH}}}{\dot{\Sigma}_\star} \frac{\int_{m_{\rm{min}}}^{m_{\rm{max}}}m_\star^{1-\delta}dm_\star}{\int_{m_{\rm{trans}}}^{m_{\rm{max}}}m_\star^{-\delta}dm_\star}\\&
\simeq 2.2 \times 10^7 \text{ years } \\& \quad \quad\times \left(\frac{S_{\rm{sBH}}}{41 \text{ pc}^{-2}}\right)  \left(\frac{\dot{\Sigma}_\star}{1.8 \times 10^{-3} \dot{\rm{M}}_{\rm{Edd}} \text{ pc}^{-2}}\right)^{-1}\,.
\end{split}
\label{eqn:tsg}
\end{equation}
In Fig.~\ref{fig:Timescales} we have plotted $t_{\rm{SG}}$ as a function of radius (blue) for each of our $\dot{M}_{\rm{out}}$ cases. For comparison we have also emphasized the expected AGN lifetime $t_{\rm{AGN}} = 10^8$ years with a horizontal gray line. 

In the three highest $\dot{M}_{\rm{out}}$ models, $t_{\rm{SG}}$ is between $10^6$ and $10^7$ years when $r > 0.1$ pc, exterior to the opacity gap. We therefore expect that this outer region will be fully supported by sBHs well within the disk lifetime. Interior to the opacity gap, $t_{\rm{SG}}$ increases sharply such that $t_{\rm{SG}} > t_{\rm{AGN}}$, suggesting that the inner disk never reaches an sBH-only supported state. We conclude that the outer disk will quickly reach a single steady-state disk solution -- not unlike the `pile-up' solution described in \citet{gilbaum2022}. The inner disk, in contrast, will continue to change over the disk lifetime as the sBH population grows.  

In the $\dot{M}_{\rm{out}} = 0.1 \dot{\rm{M}}_{\rm{Edd}}$ case, $t_{\rm{SG}} < t_{\rm{AGN}}$ exterior to $r \gtrsim 1$ pc, suggesting that a relatively small fraction of the disk will reach a steady-state in the disk's lifetime. Note, however, that our calculation of $t_{\rm{SG}}$ does not account for changes in $\dot{M}$ between the initial and final disk states. Beyond $r \gtrsim 1$ pc in the starburst-only model, $\dot{\Sigma}_{\star}$ is comparable to the higher $\dot{M}_{\rm{out}}$ cases -- the high star formation rate depletes the disk and both $\dot{M}$ and $\dot{\Sigma}_{\star}$ drop precipitously interior to $r \simeq 1$ pc. The sBH-only model, on the other hand, maintains a nearly constant $\dot{M}$ across the disk and requires auxiliary pressure support from sBHs down to $r \sim 0.1$ pc. As the outer disk evolves from the starburst-only to the sBH-only state, the mass flux through the disk will increase and is expected to ignite star formation in the disk interior. Enhanced star formation will decrease the time required for the disk to reach the sBH-only, final state.  While the change in mass flux is most pronounced in the lowest $\dot{M}_{\rm{out}}$ case, we expect it to affect star formation over time regardless of boundary conditions. 

While $t_{\rm{SG}}$ represents the time required for the disk to reach an sBH-only state, the local viscous time 
yields an estimate over which 
the gas surface density can change to adjust to the evolving number of sBHs.
We plot the viscous time in the starburst-only ($t_{\nu_{\star}}$) and sBH-only disk models ($t_{\nu_{\rm{sBH}}}$) in Fig. ~\ref{fig:Timescales}. The change in surface density between the starburst-only and sBH-only disk models is parameterized by the timescale $t_{\Sigma}$, calculated as
\begin{equation}
    t_{\Sigma} = \frac{\Sigma_\star}{\left| \Sigma_{\star} - \Sigma_{\rm{sBH}}\right|} t_{\rm{SG}}
\,,\end{equation}

that is, the surface density in the starburst-only case divided by the approximate rate at which the surface density changes as the sBH population grows. If the viscous timescale exceeds $t_{\Sigma}$, then the transition time required to reach the steady-state solution is longer than the rate at which we assume $\Sigma$ is changing. Note that, in the two highest $\dot{M}_{\rm{out}}$ cases, $t_{\Sigma} \lesssim t_{\nu_{\star}}$ and $t_{\nu_{\rm{sBH}}}$, suggesting $\Sigma$ can efficiently reach equilibrium according to the changing mass flux through the disk. But for lower $\dot{M}_{\rm{out}}$, $t_{\Sigma} > t_{\nu_{\star}}$ and $t_{\nu_{\rm{sBH}}}$, where $t_{\rm{SG}} < t_{\rm{AGN}}$, thus limiting the applicability of our model in these cases. We expect that a fully time-dependent, hydrodynamic model will be necessary to evaluate these cases, although this falls outside the scope of the current paper. 

It is useful at this point to revisit our discussion of the effect of sBH growth on our disk models. As mentioned in \S\ref{sec:implementation}, we do not take into account the growth of sBHs, setting the sBH mass to $10 \rm{M}_{\odot}$ in the models described here and in the following section. We assume, in accordance with Fig.~\ref{fig:Timescales}, that the sBHs in the outer disk will be seeded within the first ten million years of the disk lifetime. If they then undergo Eddington capped accretion for the remaining $\sim 9 \times 10^7$ years, 
both their accretion rate and mass will have increased by approximately an order of magnitude. By comparing the rates of mass flux through the disk and gas consumption by sBHs presented in Fig.~\ref{fig:steadyState}, we can see that an order of magnitude increase in accretion by sBHs is still well below $\dot{M}$. We therefore do not expect sBH growth to significantly change the evolution of the interior disk and can therefore safely neglect it.

Finally, we note that over the lifetime of the AGN, the SMBH will grow at approximately the same rate as the mass flux through the disk interior. We do not incorporate this growth into our models but can estimate a change in mass of approximately $10^6~\rm{M}_\odot$ assuming a growth rate of $0.1 ~\dot{\rm{M}}_{\rm{Edd}}$ over $10^8$ years. This amounts to an error of 25\%
in our models, which should not significantly affect our results.

\section{Disk~evolution~as~a~sequence~of~steady~states}
\label{sec:evolution}

In this section we model changes in AGN over time assuming that their evolution can be approximated as a series of steady states. This choice is justified so long as $t_{\rm{th}} \ll t_{\rm{dyn}}$ and $t_{\nu} < t_{\Sigma}$. The former assures that the disk can self regulate and, as discussed in \S\ref{subsec:Timescales}, is satisfied for reasonable disk parameters.  The latter is not satisfied for all boundary conditions, but is expected when $\dot{M}_{\rm{out}} \gtrsim 0.6 ~\dot{\rm{M}}_{\rm{Edd}}$. Nevertheless, for completeness, we analyze disks with $\dot{M}_{\rm{out}} = 1.0$, 0.6, 0.3, and 0.1 $\dot{\rm{M}}_{\rm{Edd}}$, but highlight the caveat that the disks with lower mass accretion rates  will require full, time dependent, hydrodynamic equations to confirm the results found here. 

We assume, for all models in this section, that the X-ray emission from sBHs interacts with the disk via photoionization, dust absorption and scattering, and Compton scattering. Our X-ray opacity is therefore consistent with the sBH-only models as opposed to the chimney models described in \S\ref{sec:steadystate}. Although, as discussed there, we do not expect the choice of X-ray opacity to have a significant effect on the physical disk structure, it does affect the emerging hard X-ray spectrum.

\subsection{Star formation and sBH accretion in the evolving disk}
\label{subsec:Sequence Parameters}

\begin{figure}
\centering
 \includegraphics[width=3in]{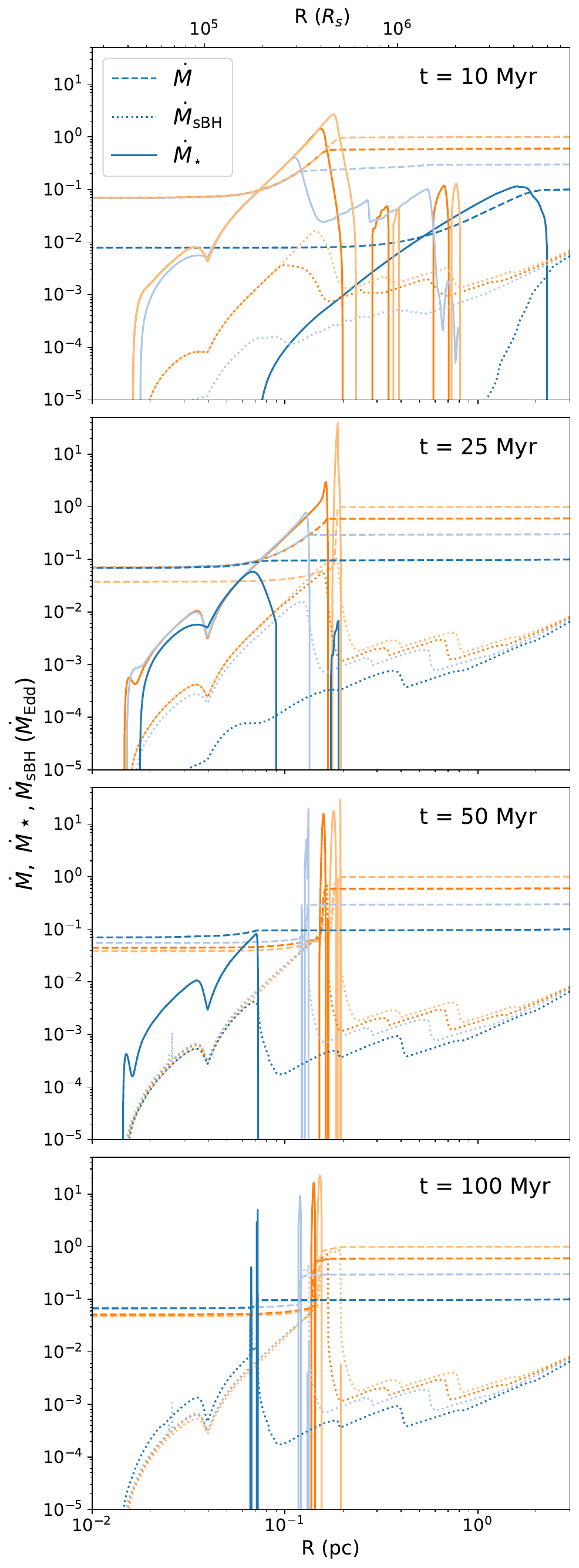}
 \caption{The total star formation rate ($\dot{M}_{\star} = 2 \pi r^2\dot{\Sigma}_{\star}$; solid line), 
 the mass flux in the background disk
 ($\dot{M}$; dashed line), and the total accretion rate onto sBHs ($\dot{M}_{\rm{sBH}} = 2 \pi r^2\dot{\Sigma}_{\rm{sBH}}$; dash-dotted line) as a function of disk radius for boundary conditions $\dot{M}_{\rm{out}} = 1.0$ (light orange), 0.6 (dark orange), 0.3 (light blue), and 0.1 (dark blue) $\dot{\rm{M}}_{\rm{Edd}}$. Moving downwards, panels illustrate $\dot{M}$ profiles at 10, 25, 50, and 100 Myr. In all models, the disks evolve toward a state in which star formation is amplified but limited to a narrow annulus. This star formation peak persists across the rest of the disk lifetime.}
    \label{fig:evolvingDisk}
\end{figure}

\begin{figure*}
\centering
 \includegraphics[width=7in]{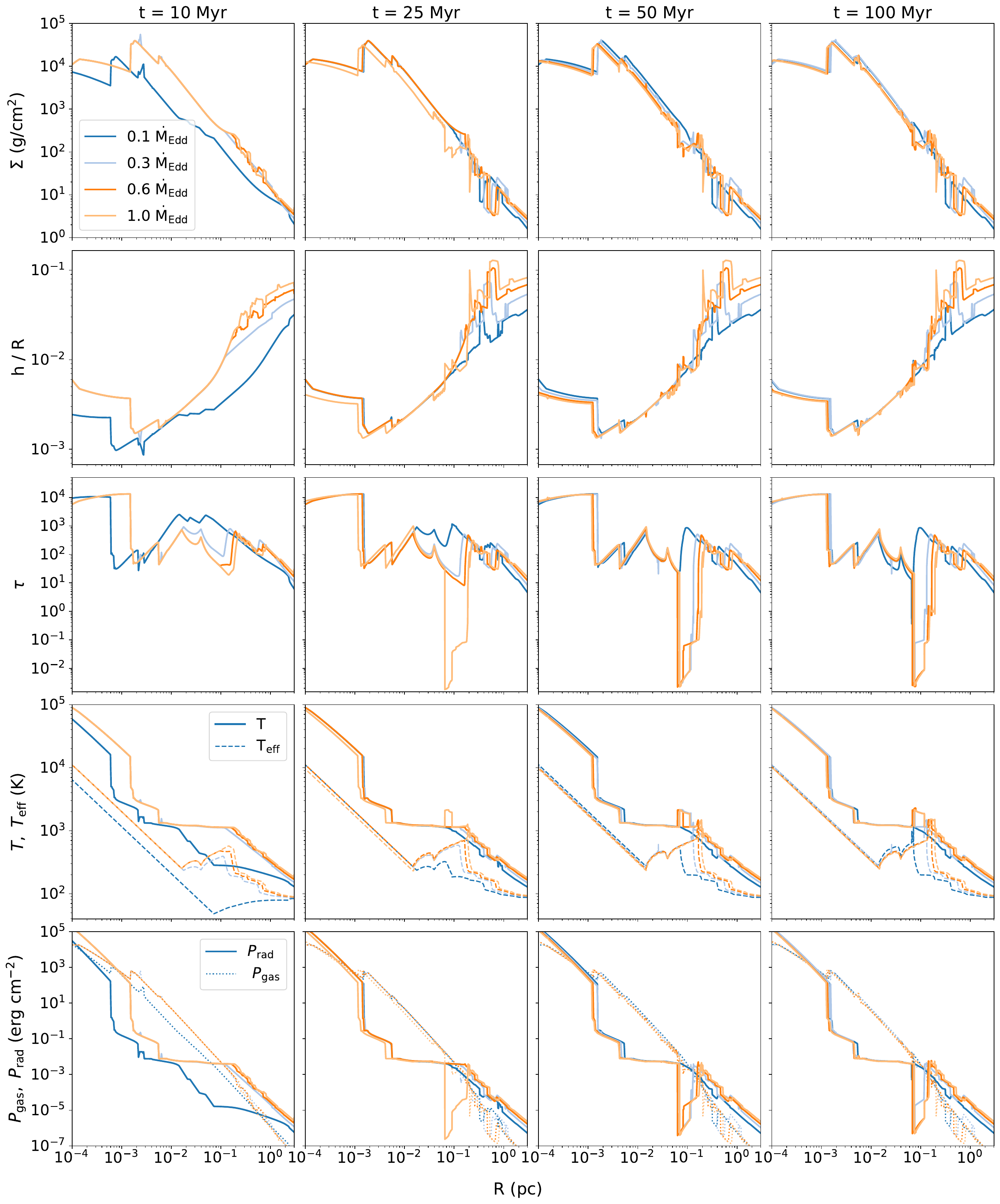}
 \caption{From top to bottom, different rows show the profiles
 of different physical quantities: surface density ($\Sigma$); ratio of scale height to radius $h/R$; optical depth ($\tau$); temperature ($T$, solid line) and effective temperature ($T_{\rm{eff}}$, dashed line); and radiation pressure ($p_{\rm{rad}}$, solid line) and gas pressure ($p_{\rm{gas}}$, dotted line). Columns from left to right
 show snapshots at 10 Myr, 25 Myr, 50 Myr, and 100 Myr. Colors are used to differentiate boundary conditions as in Fig. \ref{fig:evolvingDisk}.}
    \label{fig:evolvingDiskParams}
\end{figure*}

The dashed lines in Fig.~\ref{fig:evolvingDisk} show the mass fluxes through the disk for our four boundary conditions. Solid and dotted lines indicate the corresponding local star formation rate and sBH accretion rate, respectively. From top to bottom the four panels in Fig.~\ref{fig:evolvingDisk} provide snapshots of the disk at 10, 25, 50, and 100 Myr. 

Looking at the top panel of Fig.~\ref{fig:evolvingDisk}, we see that within the first 10 Myr, the $\dot{M}_{\rm{out}} = 1.0$ and $0.6 ~\dot{\rm{M}}_{\rm{Edd}}$ cases are largely supported by heating from sBHs at distances beyond $0.2$ pc, although rings of star formation persist at $r \simeq 0.2$ , $0.7$, and 1 pc. This is consistent with the sBH growth time ($t_{\rm{SG}}$) calculated for the outer disk in Fig. \ref{fig:Timescales},  where peaks in $t_{\rm{SG}}$ correspond to a slight shifting of the opacity away from the SMBH in the sBH-only models relative to the starburst-only models. 

\begin{figure}
\centering
 \includegraphics[width=3.2in]{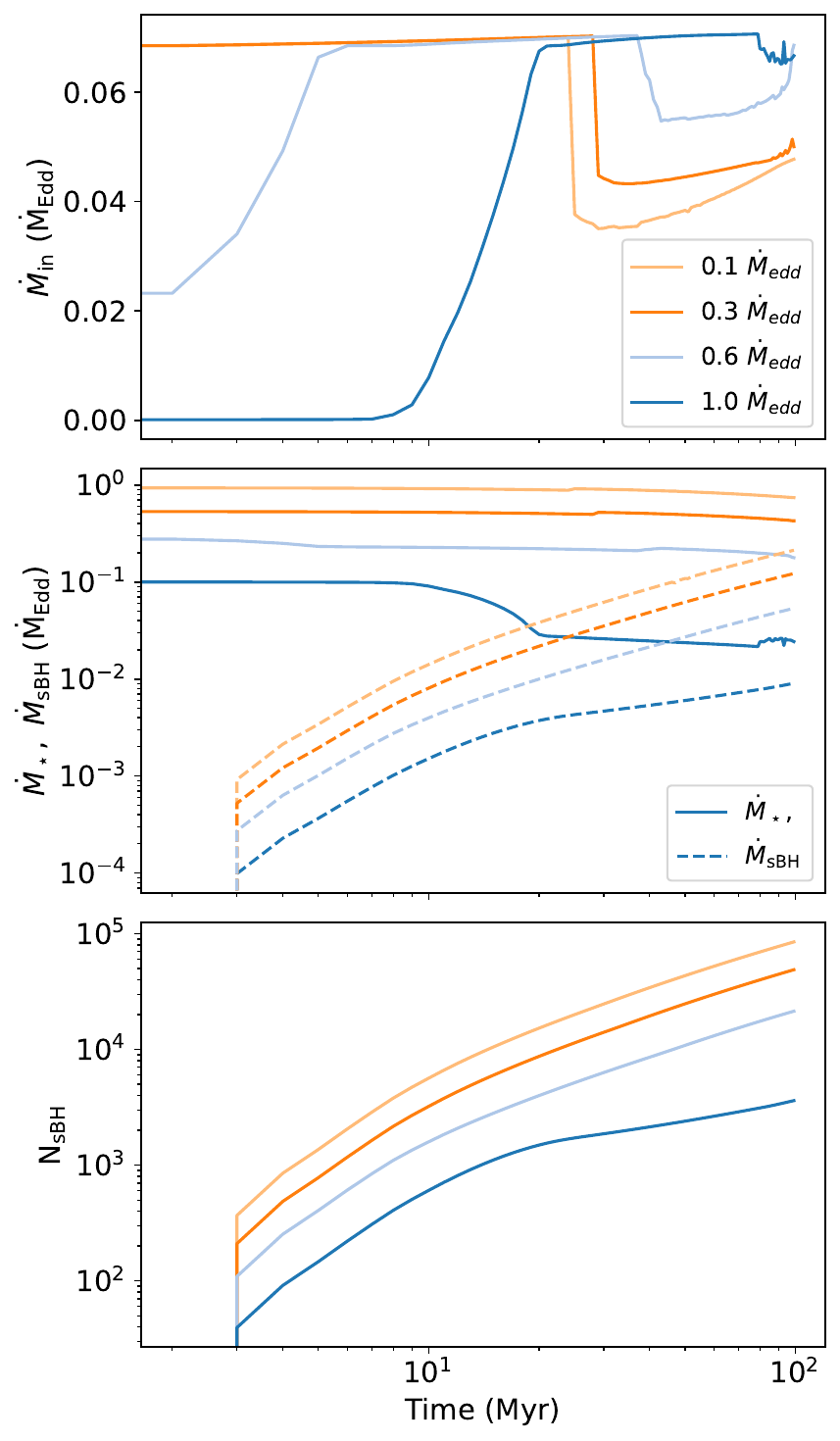}
 \caption{{\em Top panel}: Mass flux through the inner disk $\dot{M}_{\rm{in}}$ as a function of time for the four different boundary conditions: $\dot{M}_{\rm{out}} = 1~\dot{\rm{M}}_{\rm{Edd}}$ (light orange), 0.6 $\dot{\rm{M}}_{\rm{Edd}}$ (dark orange), 0.3 $\dot{\rm{M}}_{\rm{Edd}}$ (light blue), and 0.1 $\dot{\rm{M}}_{\rm{Edd}}$ (dark blue).
 {\em Middle panel}: Total mass consumption by stars i.e. $\sum_{i = 0}^n 2 \pi r_i \Delta r_i \dot{\Sigma}_\star(r_i)$ (solid line) and sBHs (dashed line) over time.
 {\em Bottom panel}: Total number of sBHs in the disk over time.}
    \label{fig:massflux_time}
\end{figure}

Star formation spans a broader portion of the disk in the $\dot{M}_{\rm{out}} = 0.3$ and $0.1 ~\dot{\rm{M}}_{\rm{Edd}}$ cases after 10 Myr. These disk models have not built up a large enough population of sBHs to support the outer disk. This is particularly clear in the lowest $\dot{M}_{\rm{out}}$ case where sBHs have yet to be seeded at distances less than 1 pc. Recall that in both of these cases, $\dot{M}_{\rm{out}} < \dot{M}_c$, and their initial starburst-only state is characterized by a high star formation rate in the outer disk and mass depletion in the inner disk. As the sBH population in the outer disk grows, star formation decreases, allowing increased mass flow and igniting star formation in the inner disk. This behavior can also be seen in the top panel of Fig. \ref{fig:massflux_time} where we plot the mass flux into the disk interior over time. Note that initially the $\dot{M}_{\rm{out}} = 0.3$ and $0.1 ~\dot{\rm{M}}_{\rm{Edd}}$ cases have $\dot{M}_{\rm{in}} = 2\times 10^{-2}$ and $1\times 10^{-4} ~\dot{\rm{M}}_{\rm{Edd}}$ respectively.  But the mass flux increases over time, reaching a maximum rate of $0.07 ~\dot{\rm{M}}_{\rm{Edd}}$ within 6~Myr in the $0.3 ~\dot{\rm{M}}_{\rm{Edd}}$ case and 20 Myr in the $0.1 ~\dot{\rm{M}}_{\rm{Edd}}$ case.

At 25 Myr, star formation is shut down beyond $\sim 0.2$ pc across all modeled disks. There are regions of the outer disk where sBH heating exceeds what necessary to maintain marginal stability. This is particularly clear in the two highest $\dot{M}_{\rm{out}}$ cases where h/r > 0.1, as shown in Fig.~\ref{fig:evolvingDiskParams}. Moreover, because we omit sBH growth and capture from the surrounding nuclear cluster, we likely underestimate the degree of excess heating. These inflated regions could experience irradiation from the disk interior, contributing to NIR variability.

25 Myr also marks a distinct transition in the behavior of the highest $\dot{M}_{\rm{out}}$ disk. As shown in the second panel from the top of Fig. \ref{fig:evolvingDisk}, the star formation rate jumps by nearly an order of magnitude at $r \simeq 0.2$ pc. The peak in star formation causes the mass flux through the disk to drop and, as a result, star formation ceases in the disk interior. This jump in star formation may be understood as a sudden switch between degenerate solutions: from a`cool,' high optical depth, radiation pressure dominated solution to a `hot,' low optical depth, gas pressure dominated solution. The jump is forced by the added heating from accreting sBHs, which increases $T_{\rm{eff}}$ and $p_{\rm{rad}}\propto \tau T_{\rm{eff}}^4$. To maintain hydrostatic equilibrium, $\tau$ and $p_{\rm{rad}}$ drop such that $p_{\rm{gas}} > p_{\rm{rad}}$. 

The gulf between the hot and cold solution parameters is more concretely illustrated in Fig. \ref{fig:evolvingDiskParams}, where we have plotted relevant disk parameters at $10$, $25$, $50$, and $100$ Myr, including $\tau$, $p_{\rm{rad}}$, and $p_{\rm{gas}}$. Note that between 10 and 25 Myr, there is a sharp drop in $\tau$ and $p_{\rm{rad}}$ at 0.1 pc in the  $\dot{M}_{\rm{out}} = \dot{M}_{\rm{Edd}}$ case, corresponding to the jump in the star formation rate. 

By 50 Myr, star formation in the $\dot{M}_{\rm{out}} = 0.6$ and $0.3 ~\dot{\rm{M}}_{\rm{Edd}}$ cases is limited to narrow rings, peaking at $r \simeq 0.16$ and $0.09$ pc respectively. And by 100 Myr all of our disk models exhibit this behavior. Although limited to a narrow ring, the cumulative mass consumption rate by stars actually increases. This is also illustrated in the top panel of Fig.~\ref{fig:massflux_time}, where $\dot{M}_{\rm{in}}$ drops steeply from $0.07~\dot{\rm{M}}_{\rm{Edd}}$ to 0.038, 0.044, 0.055, and 0.067 from highest to lowest $\dot{M}_{\rm{out}}$. The drops are slightly shifted in time, occuring at 25, 29, 40, and 83 Myr, with lower $\dot{M}_{\rm{out}}$ models taking longer to shift star formation into the hot state. In the middle panel of Fig.~\ref{fig:massflux_time} we have plotted the rates of mass consumption by stars and sBHs over time. Note that the drops in $\dot{M}_{\rm{in}}$ shown in the top panel correspond to small increases in $\dot{M}_\star$ seen in the middle panel. Mass consumption by sBHs, on the other hand, increases over time at a steady rate -- remaining subdominant to star formation over the disk lifetime. 

After 100 Myr, the star formation peak in the $\dot{M}_{\rm{out}} = \dot{M}_{\rm{Edd}}$ case has shifted slightly towards the SMBH, moving from $r \simeq 0.19$ pc to $0.15$ pc. As predicted in \S\ref{subsec:growthtime}, the timescale required for this region to become fully supported by sBHs is $\sim 10^{10}$ years. However, the enhanced rate of star formation reduces this timescale to approximately the lifetime of the disk. Here, we choose to stop evaluating our models after 100 Myr, although we expect that the star-forming rings would continue to move inward as they populated their respective regions with sBHs. However, after 100 Myr, we expect the growth of sBHs in the outer disk and in the increasingly populated inner disk to have a non-negligible effect on the disk structure. 

\begin{figure}
\centering
 \includegraphics[width=3.2in]{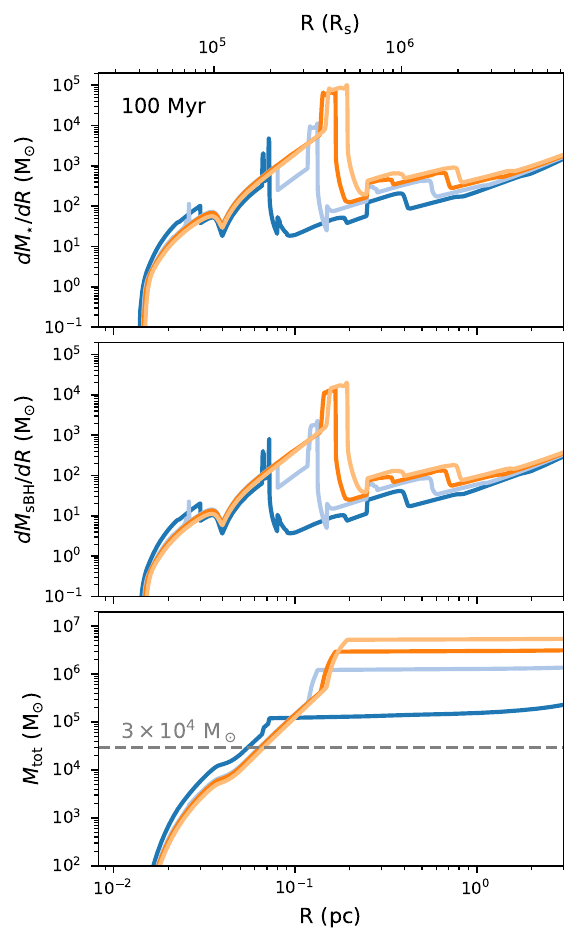}
 \caption{{\em Top Panel:} Distribution of stellar mass as a function of distance from the SMBH after 100 Myr. As in previous figures, the different colors indicate different mass supply rates. 
 {\em Middle Panel:} Distribution of sBH mass. 
 {\em Bottom panel:} Combined mass of stars and sBHs within a distance $R$ of the SMBH. The gray dashed line indicates the current maximum extended mass within $2.4\times 10^4~R_{\rm{s}}$ of the Galactic Center, as determined by \citet{2022gravity}.}
    \label{fig:mass_dist}
\end{figure}

Our models are suggestive of a nuclear disk of stars, not unlike that observed in the Galactic Center \citep{2005ghez}. Measurements of stellar orbits by \citet{2022gravity} have allowed for high-precision determination of the gravitational potential around Sgr A*. They find that the extended mass enclosed in $ 2.4\times 10^4 ~R_{\rm{s}}$ cannot exceed $3\times10^4 ~\rm{M}_\odot$. In the bottom panel of Fig.~\ref{fig:mass_dist} we plot the combined mass of stars and sBHs within a distance $R$ after 100 Myr. We find that across all models, the mass distribution is within the bounds measured by \citet{2022gravity}.  Moreover, observations of X-ray binaries indicate that there are $\simeq 10^4$ sBHs within 1 pc of Sgr A* \citep{2018hailey} and there is evidence that a population of $10^4$ sBHs within 0.1 pc is necessary to produce the quasi-thermal eccentricities of a population of stars in the Galactic center known as `S-stars'\citep{2014antonini}. However, the dynamical impact of disk loss and subsequent interactions between embedded objects has not been thoroughly investigated. Given these complicating factors, direct comparison between observed and predicted mass distributions will require extensive and careful modelling.

\subsection{Numerical resolution and convergence}
\label{subsec:resolution}

\begin{figure}
\centering
 \includegraphics[width=3.2in]{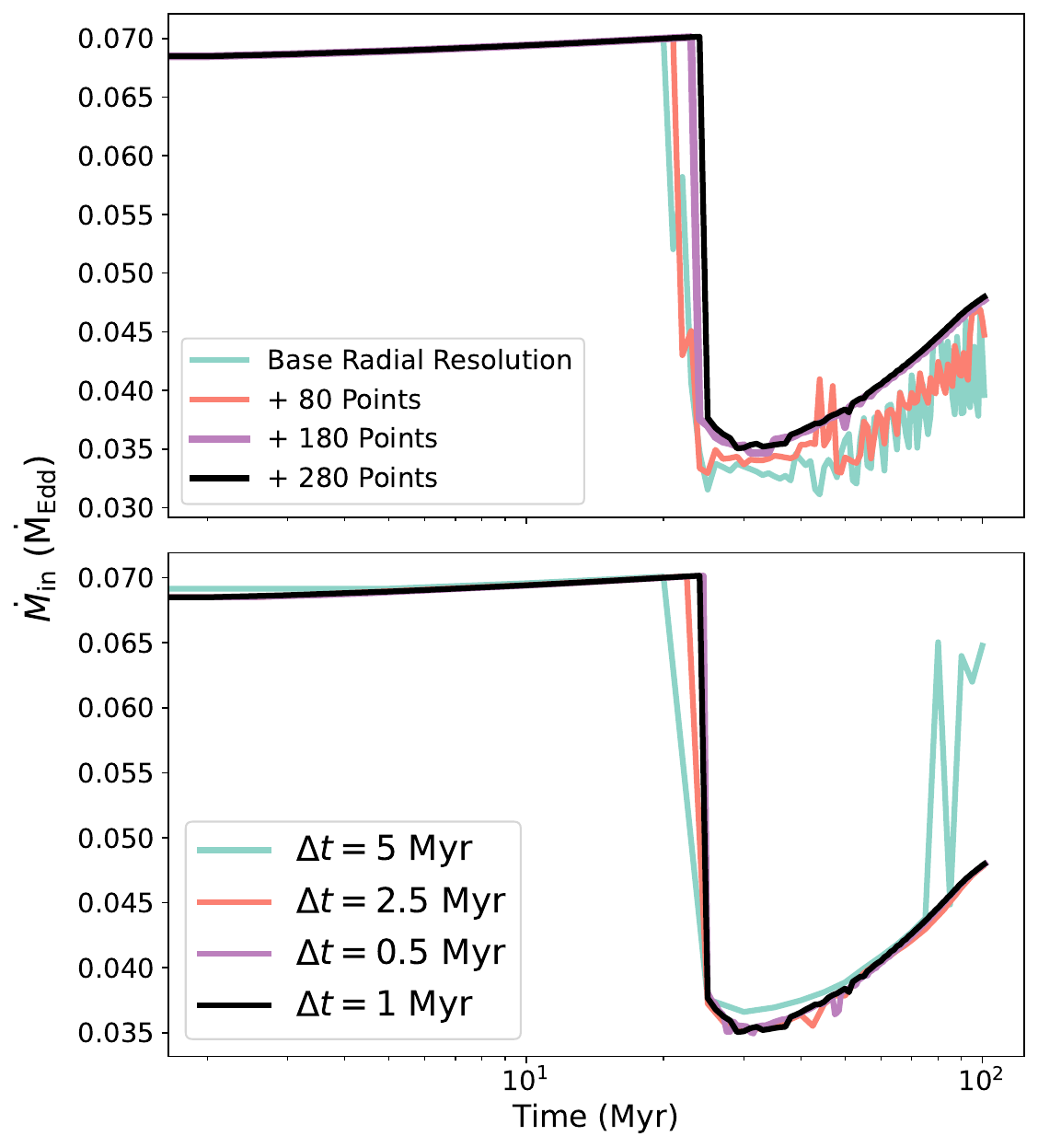}
 \caption{Evolution of $\dot{M}_{\rm{in}}$ over 100 Myr for different numerical resolutions in space and time. Black lines indicate the resolution used in the evolving models described here. All resolution tests share the boundary condition $\dot{M}_{\rm{out}} = \dot{\rm{M}}_{\rm{Edd}}$. {\em Top panel}: Spatial resolution tests including a base radial resolution of 500 points between $R_{\rm{out}} = 3$ and 0.01 pc (blue) and enhanced resolutions between 0.25 and 0.01 pc by an additional 80 (red), 180 (purple), and 280 (black) points. Models converge for resolutions exceeding 180 additional points. {\em Bottom panel}: Temporal resolution tests where $\Delta t$ is set to 5 Myr (blue), 2.5 Myr (red), 0.5 Myr (purple), and 1 Myr (black). Models converge for $\Delta t \leq 2.5$ Myr.}
    \label{fig:resolution}
\end{figure}

Our code requires us to verify convergence of our results with respect to both the
temporal and spatial resolution used. The spatial resolution is particularly critical given that, for a large part of the disks's lifetime, star formation and the resulting mass flux change occurs within a very limited radial range. We perform radial resolution tests for $\dot{M}_{\rm{out}} = \dot{\rm{M}}_{\rm{Edd}}$, assuming resolution convergence in these tests to apply for other reasonable boundary conditions. Our tests include a base resolution of 500 logarithmically spaced points in the Toomre unstable region -- between $R_{\rm{out}}$ and $0.01$ pc. The rest of the disk is less highly resolved with 304 points between 0.01 pc and $R_{\rm{in}}$. We then further enhance the resolution interior to the opacity gap -- the region where star formation persists in narrow rings -- increasing the number of logarithmically spaced radial points between $2.5 \times 10^{-5}$ and $10^{-2}$ by 400, 500, and 600 points, representing an enhancement in resolution over the base model of 80, 180, and 280 points respectively. 

In the top panel of Fig. \ref{fig:resolution} we plot $\dot{M}_{\rm{in}}$ over 100 Myr for each of our  radial resolution tests. We note that this parameter is sensitive to small changes in the star formation rate, making it a sensitive probe of resolution convergence. While all of our tests exhibit the same general behavior, $\dot{M}_{\rm{in}}$ drops slightly earlier in the two lowest resolution cases. The two lowest resolution cases also see increased fluctuation in $\dot{M}_{\rm{in}}$ after the drop. The difference between a resolution enhancement of 180 and 280 points is, however, minor, so we conclude that at these resolutions our model has converged. In the disks examined here we use the highest radial resolution from our tests, although decreasing the resolution in the opacity gap by 100 points yields comparable results. 

The bottom panel of Fig. \ref{fig:resolution} shows the results of our time resolution tests, comparing $\dot{M}_{\rm{in}}$ for models with $\Delta t = 0.5$, 1, 2.5, and 5 Myr. Our plots indicate that,  for resolutions at or below $\Delta t = 2.5$ Myr, our models converge. Even $\Delta t = 5$ Myr looks fairly smooth until until $t \simeq 70$ Myr, at which point $\dot{M}_{\rm{in}}$ increases significantly, diverging from values found at higher resolution solutions. These results suggest that our models are much less sensitive to time resolution than radial resolution, and converge for $\Delta t \leq 2.5$ Myr. 
This is reasonable given that $2.5$ Myr is the approximate time delay between star formation and evolution of the most massive stars off the MS. In our models 
we use $\Delta t = 1$ Myr.

\subsection{Emerging Spectral Energy Distributions}
\label{subsec:Sequence Spectra}

\begin{figure}
\centering
 \includegraphics[width=2.8in]{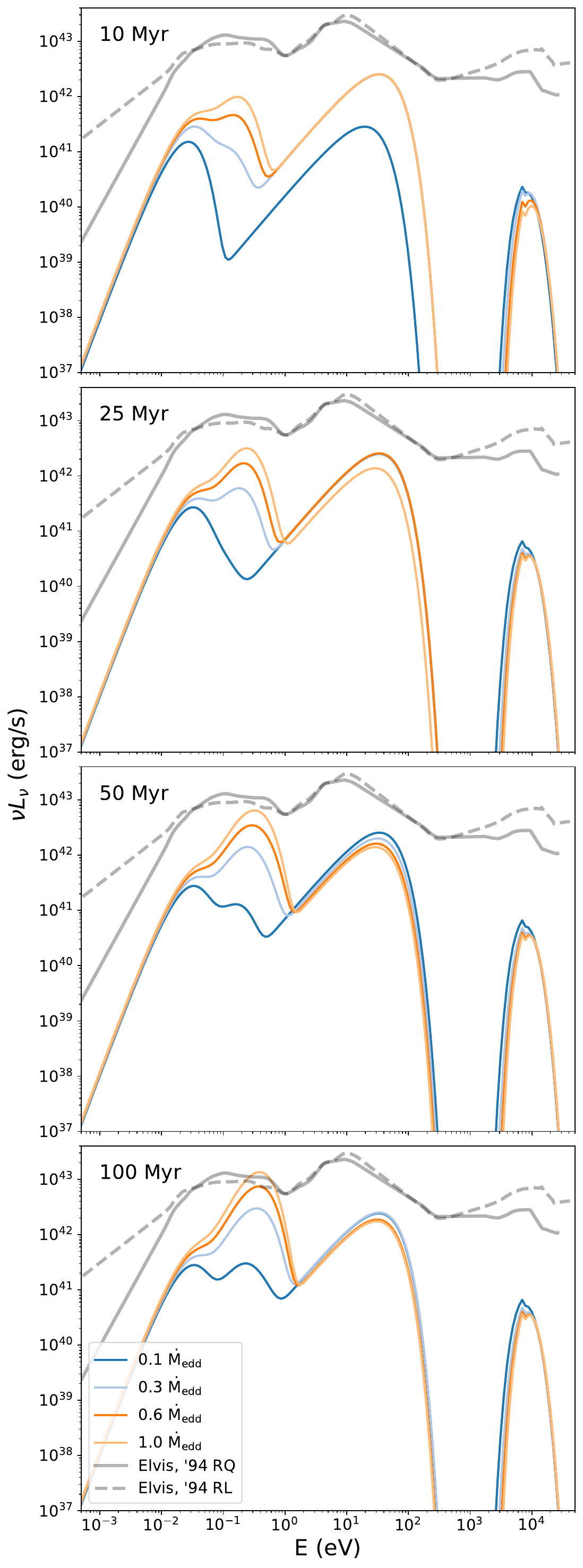}
 \caption{Spectral energy distributions (SEDs) emerging from our evolving disk models shown at 10 Myr, 25 Myr, 50 Myr, and 100 Myr (from top to bottom panel) as in the disk solutions presented in Fig.~\ref{fig:evolvingDiskParams}. The spectra shown include both the intrinsic disk emission and the escaping SED of the embedded sBH population -- visible as a high energy bump peaking at $\sim 10^4$ eV. Gray lines indicate the averaged observed emission from radio quiet (solid) and radio loud (dashed) AGN taken from \citet{1994elvis}, normalized by a factor of $5\times10^{-3}$ as in Fig. \ref{fig:spectra}. Spectral features apparent in the steady-state models (Fig. \ref{fig:spectra}) can also be recognized here, including: a mid-IR bump ($E \lesssim 0.1$ eV), a NIR bump ($0.1~\rm{eV} \lesssim E \lesssim 1$ eV), a UV-bump ($1~\rm{eV} \lesssim E \lesssim 20$ eV), and an X-ray bump that peaks at $\sim 10$ eV.}
    \label{fig:SpectraEvolving}
\end{figure}

In Fig. \ref{fig:SpectraEvolving} we show the spectra emerging from our evolving disk models. From the top-most to bottom-most panel, our spectra are taken at 10, 25, 50, and 100 Myr -- as in the disk solutions presented in Figs. \ref{fig:evolvingDisk} and \ref{fig:evolvingDiskParams}. These spectra include both the intrinsic disk flux (extending to $2 \times 10^2$ eV) and the escaped X-ray flux from embedded sBHs (peaking at $\sim$10 keV). As in Fig. \ref{fig:spectra}, the composite quasar spectra from \citet{1994elvis}, normalized by a factor of $5\times10^{-3}$, are plotted in all panels for reference.

As discussed in \S\ref{subsubsec:X-ray}, the escaped flux from sBHs comes almost exclusively from the outer disk. This population of sBHs is seeded within the first $\sim$25 Myr of the disk lifetime, after which the escaped sBH X-ray emission is constant. From smallest to largest $\dot{M}_{\rm{out}}$, the ratios of escaped sBH emission to bolometric luminosity at 25 Myr are $8.0 \times 10^{-3}$, $5.0 \times 10^{-3}$, $3.5 \times 10^{-3}$, and $3.3 \times 10^{-3}$. In general, these values represent a maximum over the AGN lifetime. At earlier times, the sBH population in the outer disk is still growing, and its emission contributes a smaller fraction to the total. At later times, the sBH emission has stabilized but thermal disk emission increases. An exception occurs in the lowest $\dot{M}_{\rm{out}}$ case, where the fractional contribution of escaped sBH emission is an order of magnitude larger at 10 Myr than at 25 Myr, as the SMBH feeding rate—-and thus the bolometric luminosity—-is significantly reduced compared to the other disk models.

For all boundary conditions, the ratio of escaped sBH emission to total disk luminosity predicted by our evolving disk models exceeds the sBH-only cases described in \S\ref{sec:steadystate} and the observational detection limit of $\sim 10^{-3} ~L_{\rm{bol}}$ estimated in \S\ref{subsubsec:X-ray}. In all cases except for the 10 Myr snapshot in the lowest $\dot{M}_{\rm{out}}$ case, the discrepancy is relatively small—-within an order of magnitude of the imposed limit—-but underscores the clear-cut constraint provided by X-ray observations. As noted earlier, a detailed numerical investigation of this constraint is beyond the scope of this work and would require a more sophisticated analysis of sBH emission beyond a multi-color blackbody spectrum. Still, the expected sBH flux, spanning 2 to 20 keV, lies well within the range detectable by \textit{NuSTAR} (3–79 keV) \citep{2013harrison}. Observational techniques, such as reverberation mapping, polarimetric analysis, and the measurement of microlensing-induced time delays \citep{2018tie}, are essential for disentangling sBH emission from the thermal emission intrinsic to the AGN disk. These methods leverage distinguishing characteristics of escaped sBH X-ray emission predicted by these models: it originates in the outer disk, is unpolarized, and varies on long timescales.

Unlike the X-ray bump, which primarily originates from sBHs in the outer disk, NIR emission reflects the number of sBHs and their contribution to disk heating interior to the opacity gap. As a result, prior to 25 Myr, the intrinsic disk spectra closely resemble the starburst-only spectra shown in Fig. \ref{fig:spectra}. At 10 Myr, the three highest $\dot{M}_{\rm{out}}$ models display broadened IR peaks, in which the mid-IR (MIR) emission ($E \lesssim 0.1$ eV) is dominant or comparable to the NIR emission ($0.1~\rm{eV} \lesssim E \lesssim 1$ eV), and both are subdominant to the UV peak ($1~\rm{eV} \lesssim E \lesssim 20$ eV). The lowest $\dot{M}_{\rm{out}}$ case does not exhibit a NIR peak at 10 Myr, but has significantly enhanced UV emission over the starburst-only model, driven by a two-order-of-magnitude increase in the mass flux reaching the disk interior.

Over time, increased mass flow to the disk interior drives enhanced star formation rates and sBH seeding, strengthening emission in the NIR. By 100 Myr the NIR bump in the three highest $\dot{M}_{\rm{out}}$ cases is the dominant feature in their respective spectra -- with peak emission nearly two orders of magnitude larger than the adjacent $\sim 1~\mu$m dip. In contrast, the \citet{1994elvis} data suggest that the depth of the $1~\mu$m dip is generally a factor of $\sim$2 smaller than the NIR peak. While significant dispersion exists in the composite spectra, with variations up to an order of magnitude cited by \citet{1994elvis}, and dominant IR peaks consistent with our data have been seen in a handful of AGN \citep{2019brown}, the degree of divergence between our results and composite spectra suggest that our fiducial model needs to be modified to accommodate a broader variety of AGN.

There are several ways the IR and amplitude can be reduced. First, as Fig. \ref{fig:SpectraEvolving} shows, if star formation ceased after $\sim$25 Myr, due to feedback or lack of fuel, then the IR bump would not grow to values well above the UV emission. Alternatively, if the outer disk size was truncated, eliminating much of the star- and BH-forming outer disks, it would similarly diminish the IR bump (as also noted  by \citealt{sirko2003}). Finally, if the angular momentum transport was more rapid than we adopted in the $\alpha$-prescription, for example due to global gravitational torques, then, for fixed boundary conditions, the surface density and the IR amplitude would be lower. Moreover, reducing the number of sBHs necessary to support the disk also suppresses the excess X-ray emission from sBHs. We examine this final dependency as well as spectral dependence on SMBH mass in the next section.

\section{Viscosity and SMBH mass dependence}
\label{sec:HighMass}

\begin{figure}
\centering
 \includegraphics[width=3.2in]{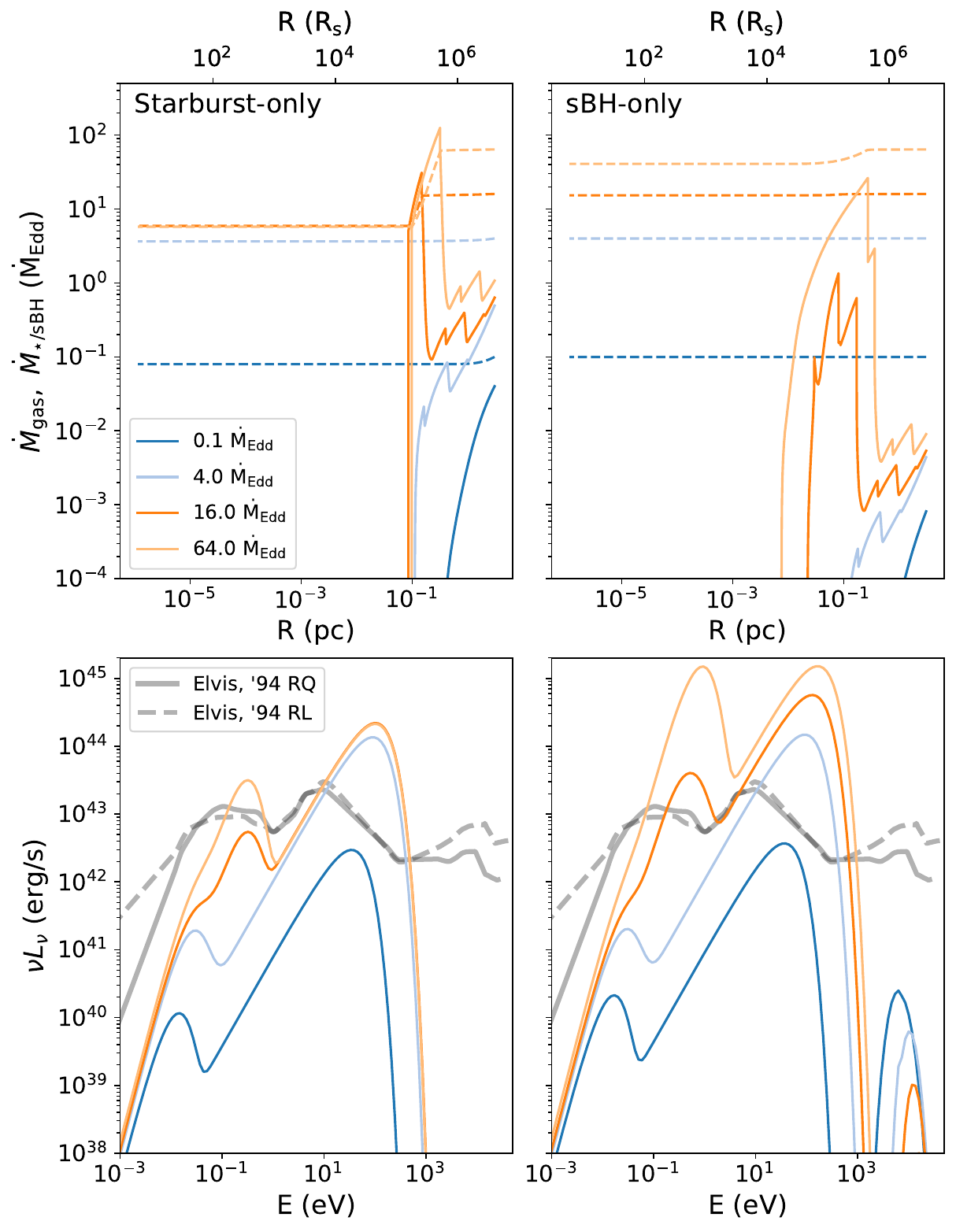}
  \caption{{\em Top panels}: Gas disk mass flux (solid lines) and mass consumption by stars (dashed lines, left panel) and sBHs (dashed lines, right panel) for single source AGN disk models with central SMBH mass of $4 \times 10^6 ~\rm{M}_{\odot}$ and assuming disk viscosity enhanced by global torques where $m = ~0.2$. Mass supply rates are distinguished by color: $1 ~\dot{\rm{M}}_{\rm{Edd}}$ (dark blue), $8 ~\dot{\rm{M}}_{\rm{Edd}}$ (light blue), $16 ~\dot{\rm{M}}_{\rm{Edd}}$ (dark orange), and $32 ~\dot{\rm{M}}_{\rm{Edd}}$ (light orange).
 {\em Bottom panels}: SEDs emerging from single source, star-only (left) and sBH-only (right), disks plotted in corresponding top panels. The \citet{1994elvis} data is plotted in gray and normalized such that the strength of UV band emission ($1 ~\rm{eV} \leq ~E \leq 10^2 ~\rm{eV}$) is $L_{\rm{UV}} = \rm{L}_{\rm{Edd}} \simeq 5.8 \times 10^{44}$ erg/s.}
    \label{fig:GT}
\end{figure}

The spectra predicted by both the single source (\S\ref{sec:steadystate}) and evolving (\S\ref{sec:evolution}) disk models diverge from typically observed AGN in three significant ways. First, the maximum mass flux reaching the central SMBH is $\lesssim ~0.1 ~\dot{\rm{M}}_{\rm{Edd}}$ with a corresponding maximum UV luminosity of $\lesssim 10^{-2} ~\rm{L}_{\rm{Edd}}$. As noted in \S\ref{subsubsec:UV}, such low accretion rates and luminosities are characteristic of the lower end of observed values and highlight the limitations of these models in reproducing the full range of AGN accretion rates and luminosities. Second, the heating required to stabilize the outer disk generally tends to raise $T_{\rm{eff}}$ and produce IR emission well in excess of a standard $\alpha$-disk. In a star-only supported disk, the resulting IR bump is roughly consistent with typical quasar spectra, but when sBH accretion feedback is included, the strength of the predicted IR bump exceeds even the most generous observational limits. Third, we predict an emission excess at $\sim$10 keV due to embedded sBHs in the outer disk, inconsistent with commonly observed AGN spectra. This detection limit puts a tight constraint on the number of sBHs expected in the disk as discussed in \S{\ref{subsubsec:X-ray}} and \S{\ref{subsec:Sequence Spectra}}.  

These discrepancies can be mitigated by increasing the disk viscosity. As noted in the previous section, higher viscosity reduces the surface density for a given $\dot{M}$, which in turn lowers the auxiliary heating necessary to stabilize the disk, damping the escaped X-ray flux from sBHs and the thermal IR emission. Additionally, lowering heating requirements lowers the corresponding rate of mass consumption by the embedded population and allows a greater mass flux to reach the disk interior.

Viscosity prescriptions beyond the standard $\alpha$-disk have been invoked by previous works to explain high rates of mass transport in luminous AGN \citep{schlosman1989, shlosman1990, goodman2003} and by TQM05 in ULIRGs. These works assume the presence of large-scale asymmetries (e.g. bars, spiral arms, or magnetic stresses), which induce rapid gas inflow. In \citet{goodman2003} and TQM05 these global torques were incorporated into steady state disk equations under the assumption that enhanced radial velocity should approach a constant fraction $m$ of the sound speed, i.e. $V_{\rm{r}} = m c_{\rm{s}}$, and Equation (\ref{eqn:Mdot}) is replaced by (\citealt{goodman2003}, TQM05):
\begin{equation}
\label{eqn:GT}
    \dot{M} = 4 \pi m \Omega h^2 \rho r \,.
\end{equation}
This simple prescription is derived from the well-known parametrization of spiral density waves \citep{binney2008}, typically applied to galactic disks. However, non-axisymmetric structures are expected to form in disks with masses exceeding $\gtrsim 0.1 ~M_{\rm{bh}}$ \citep{hopkins2011}. These structures may also be initiated and amplified by density perturbations due to ongoing star formation as well larger-scale density perturbations in the surrounding galaxy \citep{1989shlosmanb}.

To illustrate the effect of enhanced viscosity by global torques on our disk models, we incorporate Equation (\ref{eqn:GT}) into our single source, star-only and sBH-only models. As in \S\ref{sec:steadystate}, we assume Milky Way-like parameters, $M = 4\times 10^6 ~\rm{M}_{\odot}$, $\sigma = 75$ km/s, and $R_{\rm{out}} \simeq 3$ pc. Fig.~\ref{fig:GT} presents the resulting gas mass flux and mass consumption by stars (left, top panel) and sBHs (right, top panel) for $\dot{M}_{\rm{out}}$ ranging from $0.1$ to $64 ~\dot{\rm{M}}_{\rm{Edd}}$.  The corresponding emergent SEDs are shown in the bottom panels.

In the starburst-only models, the mass flux reaching the inner disk achieves a maximum value of $\simeq 6 ~\dot{\rm{M}}_{\rm{Edd}}$ when $\dot{M}_{\rm{out}} = 16$ and $64 ~\dot{\rm{M}}_{\rm{Edd}}$, nearly two orders of magnitude larger than for the disks shown in \S\ref{sec:steadystate}. The emerging spectra are bright and UV-dominant, with UV emission exceeding $10^{44}$ erg/s. 

For mass supply rates $\lesssim 6 ~\dot{\rm{M}}_{\rm{Edd}}$, star formation is suppressed, resulting in a nearly constant mass flux across the disk. The resulting spectra exhibit weak IR emission relative to the UV-bump. For $\dot{M}_{\rm{out}} = 0.1 ~\dot{\rm{M}}_{\rm{Edd}}$, the low star formation rate and constant mass flux starkly contrast with the $\alpha$-viscosity disk under identical boundary conditions, where the mass flux is reduced by three orders of magnitude. In the disk presented here, a significant fraction of the disk mass is not consumed by stars because the temperature at the outer edge of the disk is lowered to $T \simeq 50$ K. In this regime, opacity becomes less sensitive to temperature, scaling as $\kappa \propto T^{\beta}$, where $\beta < 2$. In the radiation pressure dominated outer disk, the scaling relations $\dot{\Sigma}_\star \propto \Sigma/\kappa$ and $T \propto \Sigma^{1/2}$ imply $\dot{\Sigma}_\star \propto \Sigma^{1 - \beta/2}$, that is $\dot{\Sigma}_{\star}$ scales with $\Sigma$. The temperature of the $\alpha$-viscosity disk with the same boundary conditions is $\simeq 100$ K at $R_{\rm{out}}$ and $\beta = 2$, making $\dot{\Sigma}_\star$ independent of $\Sigma$. In these cases the critical mass supply rate $\dot{M}_{\rm{c}} \simeq 2 \pi R^2 \dot{\Sigma}_{\star} \simeq 0.26 ~\dot{\rm{M}}_{\rm{Edd}}$ sets the threshold between disks that experience significant depletion and those that do not.

In the sBH-only cases, significant mass consumption occurs only for $\dot{M}_{\rm{out}} = 64 ~\dot{\rm{M}}_{\rm{Edd}}$, where the inner disk mass flux is $40 ~\dot{\rm{M}}_{\rm{Edd}}$, reduced by a factor of approximately $\sim 2/3$. In this regime, accretion heating from sBHs produces strong IR heating, comparable to the UV bump and about an order of magnitude larger than emission at the 1 $\mu$m dip. For all other cases, the mass flux through the disk remains nearly constant, so that the UV luminosities scale with $\dot{M}_{\rm{out}}$, and IR emission is a subdominant component of the spectra. The sBH contribution to the X-ray band is also significantly reduced. From lowest to highest $\dot{M}_{\rm{out}}$, the ratios of escaped sBH X-ray emission to bolometric luminosity are $2.7 \times 10^{-3}$, $1.4 \times 10^{-5}$, $4.9 \times 10^{-7}$, and $1.0 \times 10^{-8}$. As in the $\alpha$-viscosity models, this ratio maximized by lowering the mass supply rate. Only in the lowest $\dot{M}_{\rm{out}}$ case does the sBH luminosity fraction exceed $10^{-3} ~L_{\rm{bol}}$ -- the rough detection limit discussed in \S\ref{subsubsec:X-ray}. 

\begin{figure}
\centering
 \includegraphics[width=3.2in]{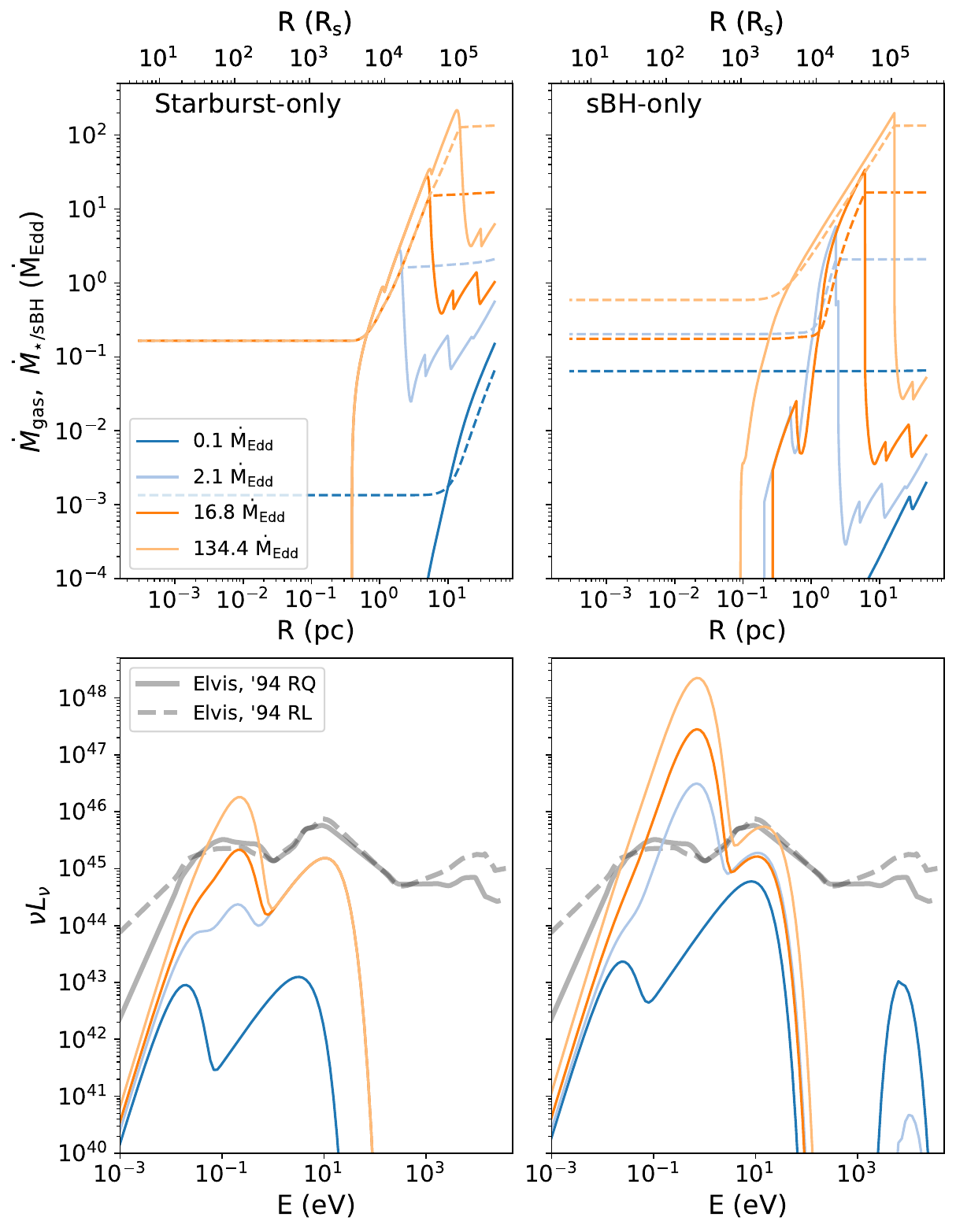}
 \caption{Assuming an SMBH mass of $M = 10^9 ~\rm{M}_\odot$, we plot mass flux and emergent SEDs for varying $\dot{M}_{\rm{out}}$, as in Figure \ref{fig:GT}. We take $\sigma = 300$ km/s, $R_{\rm{out}} = 48$ pc. In the bottom panels, we again scale the \citet{1994elvis} data so that  $L_{\rm{UV}} = \rm{L}_{\rm{Edd}} \simeq 1.4 \times 10^{47}$ erg/s.}
    \label{fig:GT2}
\end{figure}

In Fig.~\ref{fig:GT2}, we examine the effects of global torques for more commonly observed, high-mass AGN. We take $M = 10^9 ~\rm{M}_{\odot}$, $R_{\rm{out}} = 48$ pc, and $\sigma = 300$ km/s. We use mass supply rates of $0.1$, $2.1$, $16.8$ and $134.4 ~\dot{\rm{M}}_{\rm{Edd}}$, or $0.25$, $8.0$, $64$, and $512 ~\dot{M}_{\rm{c}}$, where $\dot{M}_{\rm{c}} = 6.7 ~\rm{M}_{\odot}/yr$. The resulting behavior closely resembles that of $\alpha$-disk models, where a `depleted state' emerges for sub-critical mass supply rates. Beyond this threshold, the mass flux stabilizes at just above $0.1 ~\dot{\rm{M}}_{\rm{Edd}}$ (or $\sim 4 ~\rm{M}_\odot$/yr). Constraints on mass feeding are also discussed in TQM05, where disk accretion rates are found to be similarly limited. This results in a maximum disk luminosity that is nearly an order of magnitude too faint to match the most luminous observed AGN. In the sBH-dominated models, UV emission is slightly enhanced at the highest mass supply rates, but this enhancement comes at the cost of anomalously high IR emission, creating an inconsistency with observed SEDs.

These findings suggest that an additional mechanism is required to explain the spectra of typical AGN disks, one that avoids the apparent trade-off between UV and IR emission inherent in these models. Potential candidates include enhanced viscosity in luminous AGN, driven by magnetic effects, or kinetic pressure from super-Eddington winds. Further exploration of these mechanisms could provide a more comprehensive framework for understanding the observed properties of AGN disks, particularly in the high-luminosity regime.

\section{Summary and Conclusions}
\label{sec:Conclusions}

In this work we  used a semi-analytical approach to model the dual feedback mechanisms of sBH accretion and star formation in AGN disks. We develop a set of 1D, steady-state equations that incorporate a population of embedded sBHs seeded by preceding generations of star formation. In this way, we formulate an implicitly time dependent system, through which we can approximate the evolving structure of the AGN disk and its spectral signatures. 

We assume that radiation from embedded stars and sBHs heat the disk uniformly. As in TQM05,  we expect that the UV radiation from massive stars is absorbed and scattered by dust grains which reprocess the UV emission in the IR. For this reason, the disk is always considered optically thick to UV emission. We therefore weigh the stellar mass to luminosity conversion efficiency by the UV flux of the stellar population. 

Remnant sBHs, on the other hand, are expected to emit largely in the X-ray band. These high energy photons interact with the neutral disk via ionization, dust absorption and scattering, and Compton scattering. The resulting effective X-ray optical depth can be below unity, particularly in the outer disk where surface density is low. The escaping radiation does not contribute to the disk heating and reduces the effective mass to luminosity conversion efficiency of sBHs. 

Under these assumptions and using parameters appropriate to the galactic center, we construct steady-state AGN models supported by sBHs only and compared them to star-formation supported disks. These models are analogous to the `pile-up' solutions described in \citet{gilbaum2022}, but we account for the radial depletion of gas by sBH accretion and assume efficient heat mixing in the disk. We use these models to justify our assumption of a neutral gas disk, showing that the Str\"omgren sphere for an embedded sBH does not exceed the average distance between sBHs in the Toomre-Q unstable region of the disk. Moreover, we find that ionization bubbles around the sBHs can be well approximated as a step function in the disk interior -- suggesting that AGN may maintain a two phase ionization structure at distances $\leq 0.17$ pc from the SMBH.

Beginning with a starburst-supported disk, and assuming star formation seeds the sBH population, we compute a sequence of steady state models whose structure is determined by the growing population of accreting remnants. From our starburst-only and sBH-only models we can predict a timescale over which we expect the disk to evolve. Prior to 25 Myr, our steady-state sequence models evolve consistently with these predicted timescales. But as the sBH population grows, star formation in the disk interior is enhanced. This jump in star formation depletes the mass flux in the disk interior, resulting in a narrow ring of star formation that persists over the remaining lifetime of the disk. 

We use our models to predict the effect of sBH heating on the emerging disk spectra, and identify tentative observational signatures, including enhanced emission in the NIR and the addition of a hard X-ray component. The latter comes from escaped emission from accreting sBHs and peaks at $\sim$10 keV -- overlapping an X-ray bump observed in AGN spectra commonly referred to as the `Compton hump' and attributed to reflected emission from the X-ray corona at the center of the disk \citep{2013zoghbi, 2015kara, 2021zoghbi}. The apparent absence of an emission excess below the Compton hump suggests an upper limit on the size of the embedded sBH population. Moreover, we expect emission from sBHs to differ from the Compton reflection hump in several key ways. Specifically, the escaped X-ray flux comes primarily from sBHs embedded at distances greater than 0.2 pc from the SMBH, local to the outer disk. We also expect escaped X-ray flux to vary on long timescales and be unpolarized, distinguishing it from the reflected coronal emission. Our models indicate that the relative strength and observability of this X-ray component is maximized at low mass feeding rates and during the early phases of disk evolution, when the outer disk has been populated by sBHs, while the inner disk remains primarily supported by stars. 

As demonstrated by both \citet{sirko2003} and TQM05, the inclusion of heating from star-formation in AGN disk models tends to produce IR emission well in excess of a standard steady Shakura-Sunyaev $\alpha$-disk. Additionally, mass consumed in star formation reduces the gas mass flux reaching the disk interior, limiting the SMBH feeding rate and UV luminosity to sub-Eddington rates. sBH formation in the outer disk exacerbates this behavior, because the sBHs allow a higher mass inflow rate into the inner regions of the disk, and the resulting larger surface density requires additional heating to stabilize these regions. As a result, over time, the IR bump becomes the dominant feature in our modeled spectra.  Although precise comparison with observed SEDs is beyond the scope of this simplified model, the broad results are not consistent with the UV-dominated spectra typically seen in AGN.

There are several ways the IR amplitude can be reduced. Younger or radially truncated disks would see limited sBH formation, resulting in suppressed IR emission. Alternatively, more rapid angular momentum transport via global torques reduces the auxiliary heating necessary to support the disk, thereby reducing the IR bump. We examine the latter effect by incorporating enhanced viscosity into our fiducial, Milky Way-like model as well as for a more typically observed, high mass AGN with an SMBH mass of $10^9 ~M_{\odot}$. We find that using a viscosity prescription that incorporates global torques alleviates the IR strength, X-ray excess, and UV faintness for our fiducial, low-mass AGN models. However, for higher-mass AGN, the anomalously high IR and faint UV emission persist, suggesting that our models are insufficient to explain the spectral shapes of luminous AGN. Resolving this will likely require the inclusion of additional physics, such as further enhanced viscosity driven by magnetic effects or non-thermal sources of pressure support including super-Eddington winds from embedded stars and sBHs. 

While beyond the scope of the present paper, future work will aim to quantify these constraints. Any such analysis must also account for the contribution from circumnuclear tori to AGN IR emission, as inferred from time-lags between IR and UV variability. The upcoming GRAVITY+ upgrade offers a promising avenue to disentangle the fraction of NIR emission intrinsic to AGN disks. GRAVITY has already enabled precise localization of NIR emission, as demonstrated by observations of NGC 1068, which revealed a disk-like structure rather than the traditionally assumed torus-like morphology \citep{20200pfuhl}. The GRAVITY+ upgrade will improve resolution by a factor of $\sim$2, expanding the number of AGN whose NIR and broad-line region (BLR) emission can be imaged and improving imaging of these regions in previously observed AGN. These advancements will enable a more detailed characterization of the spatial distribution of NIR emission, placing additional constraints on models involving heating by embedded sBHs.

As our paper was being finalized for submission, we became aware of a related preprint by \citet{Zhou+2024}.  While their modeling approach is quite different, they discuss the modified emission from AGN disks due to the presence of embedded sBHs.  Focusing on the optical spectrum near 5000 \AA, they find that embedded sBHs increase $T_{\rm{eff}}$ in the outer disk and that the spatially extended heating from sBHs may be resolved in microlensing observations. These results are consistent with what we found, although our model includes many additional physical ingredients, and addresses the time-evolution of the system.

In this work we focus on the contributions of star formation and consequent, in situ sBH accretion to stabilize AGN disks. To enhance the clarity of our results we have chosen to simplify our model in a number of ways. First, we have assumed the disk evolution can be treated as a sequence of steady states, and we neglect to account for the full hydrodynamic behavior of the disk. We expect this simplification to have the greatest impact on models with $\dot{M}_{\rm{out}} < \dot{M}_{\rm{crit}} \simeq 0.13 ~\dot{\rm{M}}_{\rm{Edd}}$, in which the mass flux through the inner disk changes by several orders of magnitude. Second, we have neglected to incorporate the dynamical processes influencing the size and location of sBHs and stars in the disk. These processes include: (1) angular momentum exchange between embedded objects and the gas disk, resulting in migration \citep{bellovary2016, secunda2019, secunda2020,tagawa2020}. (2) Capture of stars and sBHs from the surrounding nuclear cluster (e.g. \citealt{bartos2017}). (3) Dynamical interaction between embedded objects including scatterings and mergers. We expect to address both the hydrodynamical evolution of these disks and dynamical evolution of the stellar and sBH population embedded in them in future work.

Finally, we do not account for the growth of the sBHs or stars in these models. We show that the former will not change the qualitative conclusions drawn here, although we expect sBH growth to affect disk structure for AGN lifetimes exceeding 100 Myr. The latter, on the other hand, remains an interesting and open question. Recent work has suggested that stars embedded in AGN disks may evolve to be  
very massive, or perhaps even "immortal",
replenishing hydrogen gas via accretion at the same rate it is burned in the core \citep{2021cantiello, Jermyn2021rot,2022jermyn}. As discussed in \citet{2024chen}, such a population would diminish the effects of remnants on disk structure and evolution. On the other hand, super solar abundances observed in AGN offer strong evidence in support of ongoing stellar evolution \citep{2023huang}. To account for this apparent discrepancy between theory and observation \citet{2023alidib} suggest that stellar lifetimes in AGN may be limited by gap formation in the disk, He accumulation in stars, and suppression of mixing by the radiative layer in stellar interiors. According to this `Stellar Evolution and Pollution in AGN Disks' (SEPAD) model, star formation and evolution off the MS would lead to sufficient levels of heavy element disk pollution to account for observed abundances. Still, these models suggest a stellar IMF that may diverge significantly from field stars and vary based on disk location. This alternative stellar evolution model, and the remnant population it would create, offers another interesting avenue for future work.

To summarize, we find that incorporating sBH remnants into AGN disk models has significant impact on the disk structure, evolution and corresponding spectra. We emphasize that star formation in AGN disks persists across the disk lifetime, although its radial extent becomes increasingly limited over time, confined to a narrow band at the edge of the opacity gap or $\simeq 0.2$ pc for parameters appropriate to the galactic center, considered here. These processes predict a unique spectral AGN signature, including enhanced IR emission and an additional contribution to the X-ray band attributed to escaped flux from accreting sBHs. Neither of these features are commonly observed in AGN, but highlight observational constraints on AGN models that include auxiliary heating in order to stabilize the outer disk. This feature is likely reconcilable with observations once additional physics is incorporated into our models, which we defer to dedicated follow up study.

\medskip
\section*{Acknowledgements}
 We thank Semih Tuna, Erin Kara, Yuri Levin, Brian Metzger, Andrei Beloborodov, as well as an anonymous referee, whose thoughtful insights significantly improved this work. M.E.M.~has been supported by \textit{GFSD}. 
 ZH acknowledges financial support from NASA grants 80NSSC22K0822 and 80NSSC24K0440 and NSF grant AST-2006176. 
 R.P. gratefully acknowledges support by NSF award AST-2006839. 
 H.T. was supported by the National Key R\&D Program of China (Grant No. 2021YFC2203002). 

{\em Software:} pandas \citep{ mckinney-proc-scipy-2010}, IPython \citep{PER-GRA:2007}, matplotlib \citep{Hunter:2007}, scipy \citep{scipy}, numpy \citep{oliphant-2006-guide}, Jupyter \citep{Kluyver:2016aa}

\section*{Data Availability}
The data underlying this article will be shared on reasonable request to the corresponding author.

\bibliographystyle{mnras}
\bibliography{refs}

\end{document}